\documentclass[12pt,preprint]{aastex}

\newcommand{\kms}{$\;$km s$^{-1}$}
\newcommand{\msun}{M_{\odot}}

\begin{document}
 
\title{Time-Series Photometry of M67: W UMa Systems, Blue
Stragglers, and Related Systems}

\author{Eric L. Sandquist}
\affil{San Diego State University,Department of Astronomy,San Diego, CA 92182}
\email{erics@mintaka.sdsu.edu}

\author{Matthew D. Shetrone}
\affil{University of Texas/McDonald Observatory,
P.O. Box 1337, Fort Davis, Texas 79734}
\email{shetrone@astro.as.utexas.edu}

\begin{abstract}
We present an analysis of over 2200 $V$ images taken on 14 nights at
the Mt. Laguna 1 m telescope of the open cluster M67. Our observations
overlap but extend beyond the field analyzed by Gilliland et
al. (1991), and complement data recently published by van den Berg et
al. (2002) and Stassun et al. (2002). We show variability in the light
curves of all 4 of the known W UMa variables on timescales ranging
from a day to decades (for AH Cnc).  We have modeled the light curve
of AH Cnc, and the total eclipses allow us to determine $q =
0.16^{+0.03}_{-0.02}$ and $i = 86 \degr ^{+4}_{-8}$.  The position of
this system near the turnoff of M67 makes it useful for constraining
the turnoff mass for the cluster. We have also detected two unusual
features in the light curve of AH Cnc that may be caused by
prominences. We have also monitored cluster blue stragglers for
variability, and we present evidence hinting at low level variations
in the stragglers S752, S968, and S1263, and we place limits on the
variability of a number of other cluster blue stragglers.
Finally, we provide photometry of the sub-subgiant branch star
S1063 showing variability on timescales similar to
the orbital period, while the ``red straggler'' S1040 shows evidence
of an unexplained drop in brightness at phases corresponding to the
passage of the white dwarf in front of the giant.
\end{abstract}


\section{Introduction}

Thanks to high stellar densities and small velocity dispersions,
stellar clusters stand out as environments with high frequencies of
strong gravitational interactions between stars. In the cores of the
densest globular clusters, collisions of single stars may occur
relatively frequently (Hills \& Day 1976). However, even for less
active environments like open clusters, strong interactions between
binary stars probably play an important role in modifying the orbital
parameters of the binaries and facilitating stellar collisions (Hurley
et al. 2001; Portegies Zwart et al. 2001). Current thinking identifies
blue stragglers as one of the likely products of stellar interactions,
so that an in-depth understanding of these stars could lead to a
better understanding of the overall importance of environmental
effects within stellar clusters (see Bailyn 1995 for a review).
 
As a byproduct of our project to monitor the partially-eclipsing blue
straggler S1082 (designation from Sanders 1977), we made extensive
observations of the fields around the core of the open cluster M67. In
order to get complete coverage of the light curve of S1082 (period
1.0677978 d) observations over several nights needed to be made. After
it was discovered that the light curve was variable on timescales of a
month or less, our observational dataset was expanded further. The
observations of that system are presented and analyzed in a separate
paper (Sandquist et al. 2003). However, in combination with the large
archive of radial velocity measurements for M67 stars (e.g. Milone \&
Latham 1994), photometric studies of blue stragglers can provide us
with important clues to their current states, and may also lead to an
understanding of the evolutionary route they followed before becoming
identifiable as blue stragglers. W UMa variables are one class of
binary star that can produce blue stragglers after angular momentum
losses cause the two stars to coalesce.

M67 is a somewhat difficult target for comprehensive variability
studies because of its large angular size, but several groups have
presented results on the cluster. For example, Gilliland et al. (1991)
combined relatively deep observations from several different
observatories in a study of the core of the cluster. They
serendipitously discovered two W UMa variables (S1036 and III-79), two
blue stragglers with low amplitude $\delta$ Scuti pulsations (S1280
and S1284), and evidence of longer period variations in other
stars. Their field was relatively small, however, so that it was
nearly certain that other variables would be found.  Recently van den
Berg et al. (2002, hereafter vSVM) and Stassun et al. (2002, hereafter
SvMV) presented photometric studies of variables in a larger field in
M67. Their studies were initiated to look for variability among the
X-ray sources within the cluster. Several of the known X-ray sources
were shown to be low-amplitude variables, and photometry was presented
for an additional W UMa variable (S757).

The present study complements these photometric studies in several
ways.  In a number of instances we are able to present better
delineated light curves than vSVM and SvMV because we took shorter
exposures and focused on observations in a single filter band. In
addition, by combining our results with those of the other studies we
are able to build a better picture of the variability of the light
curves themselves on timescales ranging from days to decades for some
of the better known variables.  In \S 2 we briefly describe the
observations, in \S 3 we discuss the reduction of the photometry, and
in \S 4 we discuss the variables.
 
\section{Observations}

All of the photometry for this study was taken at the 1 m telescope at
the Mt. Laguna Observatory using a $2048 \times 2048$ CCD on 19 nights
between December 2000 and March 2002. The nights of observations are
given in Table~\ref{obs}. The photometry was primarily in $V$ band
with typical exposure times of 20 s (ranging between 15 and 60 s
depending partly on atmospheric transparency) to optimize the counts
for the variable S1082. Exposures were usually separated by about 2.5
minutes due to a relatively long readout time for the CCD.

Guiding jitter and poor air flow at the site of the telescope
typically restricts image quality to greater than 4 pixels (1.6
arcsec) in the best conditions.  Observing conditions for the nights
varied greatly, reaching FWHM of 13 pixels for some of the worst
frames. The relative sparseness of the cluster worked in our favor
though because decent photometry could still be done for many of the
stars in the field.

Several hundred columns of the chip developed charge transfer problems
during the December 2000 run. Because of this, measurements of stars
that fell near these columns were eliminated. The remainder of the
chip was not noticeably affected by this problem. In the January and
February 2001 runs this problem was almost entirely corrected. The CCD
was replaced for the March 2001 run and a two-amplifier readout was
employed, which doubled the duty cycle for our observations. For the
remainder of the observations the replacement CCD was used with a
one-amplifier readout.

\section{Analysis}

\subsection{Photometric Reduction}

The object frames were reduced in usual fashion, using overscan
subtraction, bias frames, and flat fields (usually twilight flats,
with the exception of Dec. 11/12, 2000, Feb. 17/18 and 18/19, 2001,
for which dome flats had to be used).  We chose to rely on aperture
photometry for this study. In the analysis we used the
IRAF\footnote{IRAF (Image Reduction and Analysis Facility) is
distributed by National Optical Astronomy Observatories, which are
operated by Association of Universities for Research in Astronomy,
Inc., under contract with the National Science Foundation.}  tasks
DAOFIND and PHOT from the APPHOT package. Curve-of-growth analysis was
conducted using the IRAF tool DIGIPHOT.PHOTCAL.MKAPFILE in order to
bring all photometric measurements to the same total aperture
size. The general procedure is discussed in Stetson (1990).
Individual nights were run separately through the curve-of-growth
analysis.

In order to improve the accuracy of the relative photometry for the
light curves, we used an ensemble photometry method similar to that
described by Honeycutt (1992) in order to get a simultaneous solution
for median magnitudes of all stars and relative zero-points for all
image frames. Our implementation is described in more detail in
Sandquist et al. (2003). The solution was improved iteratively until
none of the frame zero-points or median star magnitudes changed by
more than 0.0005 mag between iteration steps.  We allowed for the
possibility that magnitude residuals could be a function of star
position on the CCD for each frame by fitting second-order polynomials
to the residuals in the $x$ and $y$ directions, and subtracting these
fits during the solution iteration. Occasionally these corrections
amounted to a few centimag.

In reducing the data, we paid close attention to possible correlations
of residuals with external variables like seeing, airmass, sky
intensity, exposure time, and the photometric zeropoint of the frame
(see Gilliland et al. 1991 for a detailed discussion of the reasons).
During the photometric reductions, we discovered that the photometry
of a handful of the bluest stars in the sample showed significant
variations of approximately 0.02 mag from night to night.  On further
investigation, we found that for several of these stars the
photometric variations around the median values were definitely
correlated. Although we have been unable to determine the exact cause
of this problem, we were able to verify that the correlations in the
variation were only detectable in stars with $\bv \le 0.3$. In our
sample this affected the blue stragglers S752, 968, 977, 1066, 1263,
and 1267. None of these stars had strong variations, but in order to
try to look for low amplitude changes we computed a running weighted
average of the residuals derived from all of the stars observed on a
frame and on up to 6 other frames closest in time. The correction
was subtracted from the brightness measurement for each star. The
results of this analysis will be discussed in more detail in \S
\ref{otherv}.

\subsection{Variability Analysis}\label{vari}

\begin{figure}
\plotone{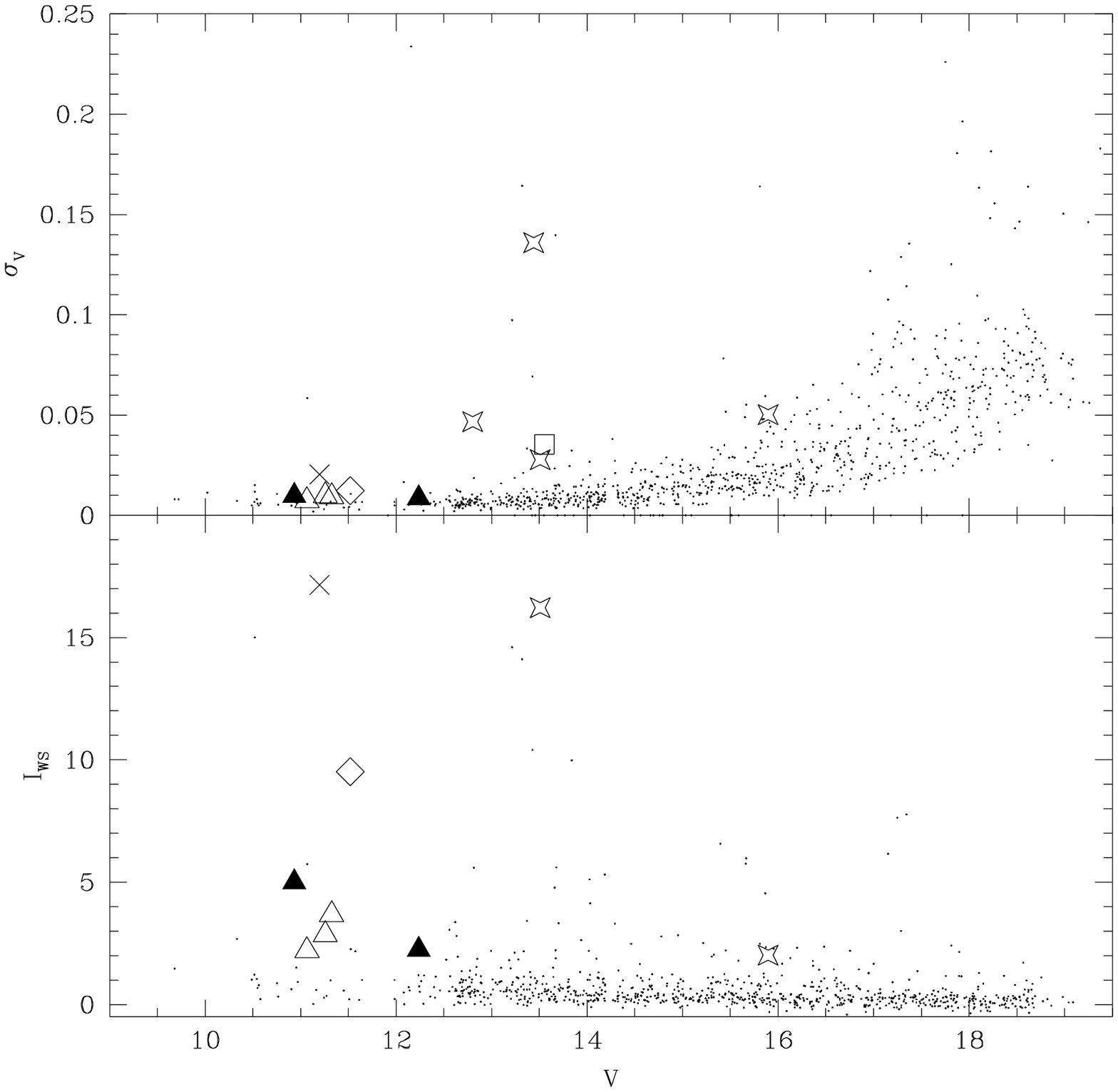}
\caption{Variability indicators $\sigma_{V}$ (rms scatter in $V$
measurements) and Welch-Stetson index $I_{WS}$ versus magnitude.
Measurements taken under poorest seeing conditions (FWHM $\ga
2\farcs5$) have been eliminated from the calculations of the plotted
indices. The meaning of the symbols is given in Fig. 2. Note that the
W UMa variables S1036 and S1282, and the sub-subgiant branch star
S1063 are off the top of the plot in the lower panel.\label{var}}
\end{figure}

To judge the significance of observed photometric variations, we used
two different measures: the rms variations about the median magnitude
$\sigma_{V}$ and the Welch-Stetson index $I_{WS}$ (Welch \& Stetson
1993). $I_{WS}$ measures the degree of correlation of pairs of
brightness measurements, and is generally the more sensitive method of
the two for detecting variability. This index gives high scores for
previously known variables, with the $\delta$ Scuti star S1280 (a
variable of low-amplitude) having the lowest score ($I_{WS} =
2.55$). In Fig.~\ref{var}, we plot both measures as a function of
magnitude for the stars in our samples to provide the reader with a
means of judging the significance of the variations discussed
below. This is complicated by the fact that seeing variations could
result in correlated residuals for a star if a nearby star of
comparable brightness contributed light to the aperture used to
photometer the star. The variability indices plotted in Fig.~\ref{var}
were calculated for measurements with the best seeing (FWHM $\la
2\farcs5$) in order to reduce this confusion, although some of the
stars with high scores are still affected by this effect. However, for
the objects discussed in this paper, we verified that there is no
correlation between the photometric residuals (defined as observed
minus median magnitude) and seeing.

We used two different techniques to determine periods for the
variables we observed: the Lafler-Kinman statistic \citep[hereafter
L-K]{lk}, and the Lomb-Scargle \citep[hereafter L-S]{lsc} periodogram.
These two statistical methods measure slightly different
characteristics of trial light curves. The L-K statistic measures the
quality of a light curve for a given trial period using the sum of the
differences in magnitude between observations made at adjacent
phases. Variations in the overall brightness of the system on
different orbits can cause problems with this test, however. The L-S
periodogram is basically a harmonic decomposition of the observations.

\subsection{Light Curve Analysis}\label{lcanal}

For the systems with the most stable light curves, we modeled the
light curve using the program NIGHTFALL\footnote{See
http://www.lsw.uni-heidelberg.de/$\sim$rwichman/Nightfall.html for the
program and a user manual (Wichmann 1998)}, which includes model
atmospheres (Hauschildt et al. 1999) and physical effects such as
detailed reflection (which is important for close binaries).  Detailed
fitting of the eclipsing binaries requires some care in choosing the
prescription for limb-darkening. Significant systematic differences in
parameters such as mass ratio and inclination can result from this
choice. We have chosen to use a two-parameter square-root law
primarily because the use of linear or quadratic limb darkening laws
resulted in significantly higher $\chi ^2$ values due to poor fits to
the eclipses for our variables. Our choice is supported by
comparisons between predictions from model atmospheres and best-fit
limb-darkening laws \citep{vh1, diaz95} that indicate that square-root
or logarithmic laws are to be preferred for stars with surface
temperatures near those of our W UMa variables.

\section{Light Curves}\label{analysis}

\begin{figure}
\plotone{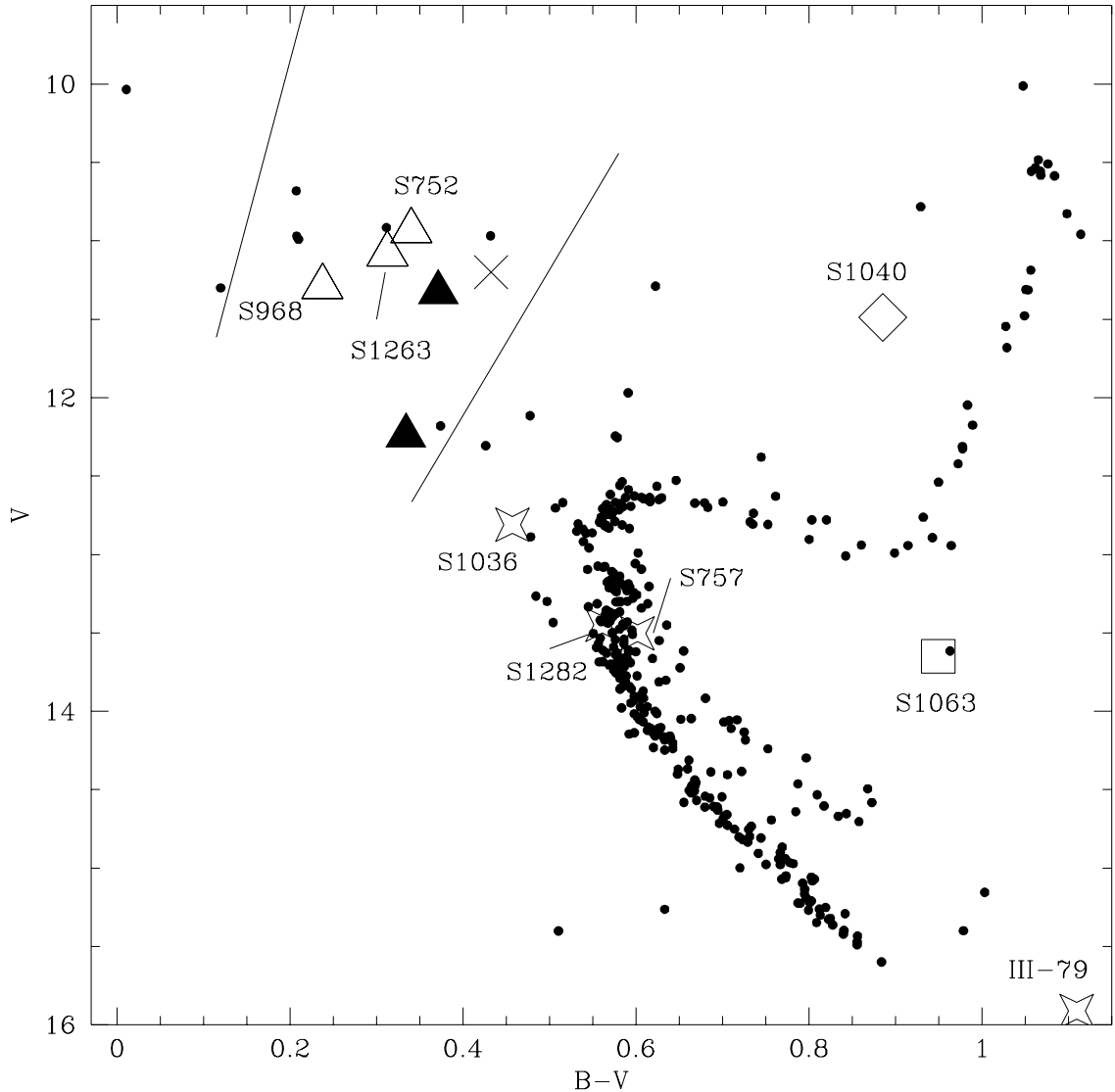}
\caption{The turnoff region (including blue stragglers) of the
color-magnitude diagram of M67 from the data of \citet{fan96} and
cleaned of probable non-members using the proper motions of
\citet{sanders77}.  The approximate boundaries of the $\delta$ Scuti
instability strip are shown with solid lines. The labelled open
symbols represent systems discussed in the text. The observed W UMa
contact binaries are marked as open stars, previously known $\delta$
Scuti pulsators are marked as filled triangles, possible low amplitude
variables are open triangles.  The sub-subgiant branch star S1063 is
an open square (with the related system S1113 a dot in the corner of
the square), the ``red straggler'' S1040 is an open diamond, and the
RS CVn system S1082 is marked with a $\times$.\label{fig1}}
\end{figure}

In this section, we discuss the important features of the light curves
of the variable stars we detected. Unless otherwise listed, the ID
numbers are from Sanders (1977). A color-magnitude diagram for the 
systems discussed below is shown in Fig.~\ref{fig1}.

\subsection{W UMa Contact Binaries}
 
Four W UMa binaries are currently known to be members of M67. We
present observations of all of these systems below, and we list their
properties in Table~\ref{wuma}.

\begin{figure}
\plotone{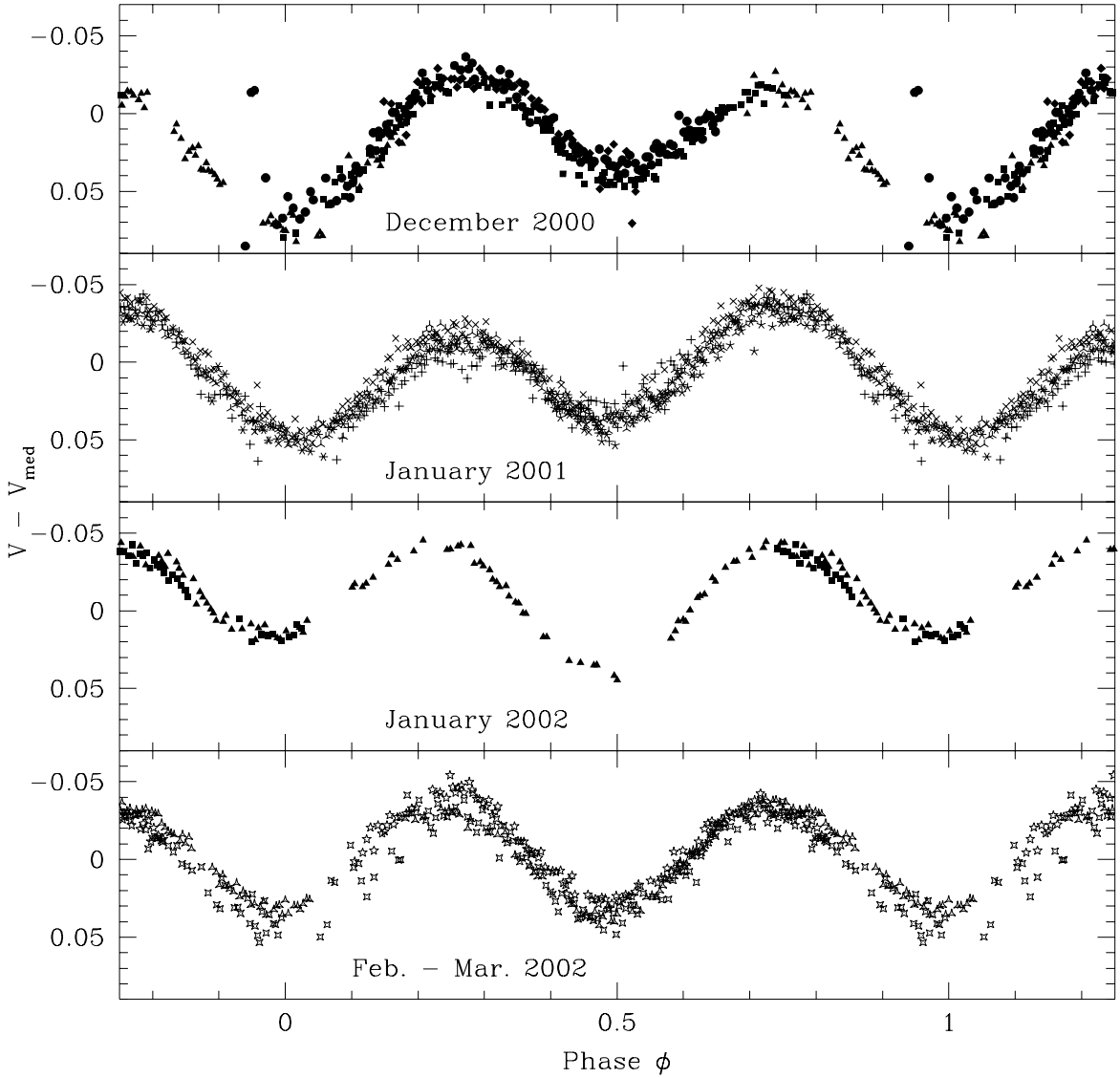}
\caption{Phased $V$ light curves for the contact binary S757,
separated by month of observation. The median $V$ magnitude used was
for the complete dataset. Zero phase was chosen to be the photometric
minimum.  Different symbols correspond to different nights of
observation (see Fig.~\ref{s757.2} for identifications).\label{s757}}
\end{figure}

{\bf S757:} This star was slightly outside the field analyzed by
Gilliland et al. (1991), but was identified as a probable variable
star by both Nissen et al. (1986) and Rajamohan et al. (1988). SvMS
discuss their identification of this system as a W UMa variable based
on a likely period of 0.3600 d and color $(B-V) = 0.61$. They
presented a partial light curve generated from two runs in 1998 and
2000 that indicate maxima and minima of roughly equal brightness. We
independently identified the system, and are able to refine the period
to $P = 0.35967 \pm 0.00002$ d from two seasons of data. In this case
we found that the best period from the L-K method was in a smaller
peak in the L-S periodogram.  The amplitude of the light curve in $V$
is approximately 0.09 mag, although this does vary from month to
month. Regardless, the small amplitude indicates that the system has a
fairly low inclination ($\sim 30\degr$).  Our data phased to this best
period are shown in Fig.~\ref{s757}.

\begin{figure}
\plotone{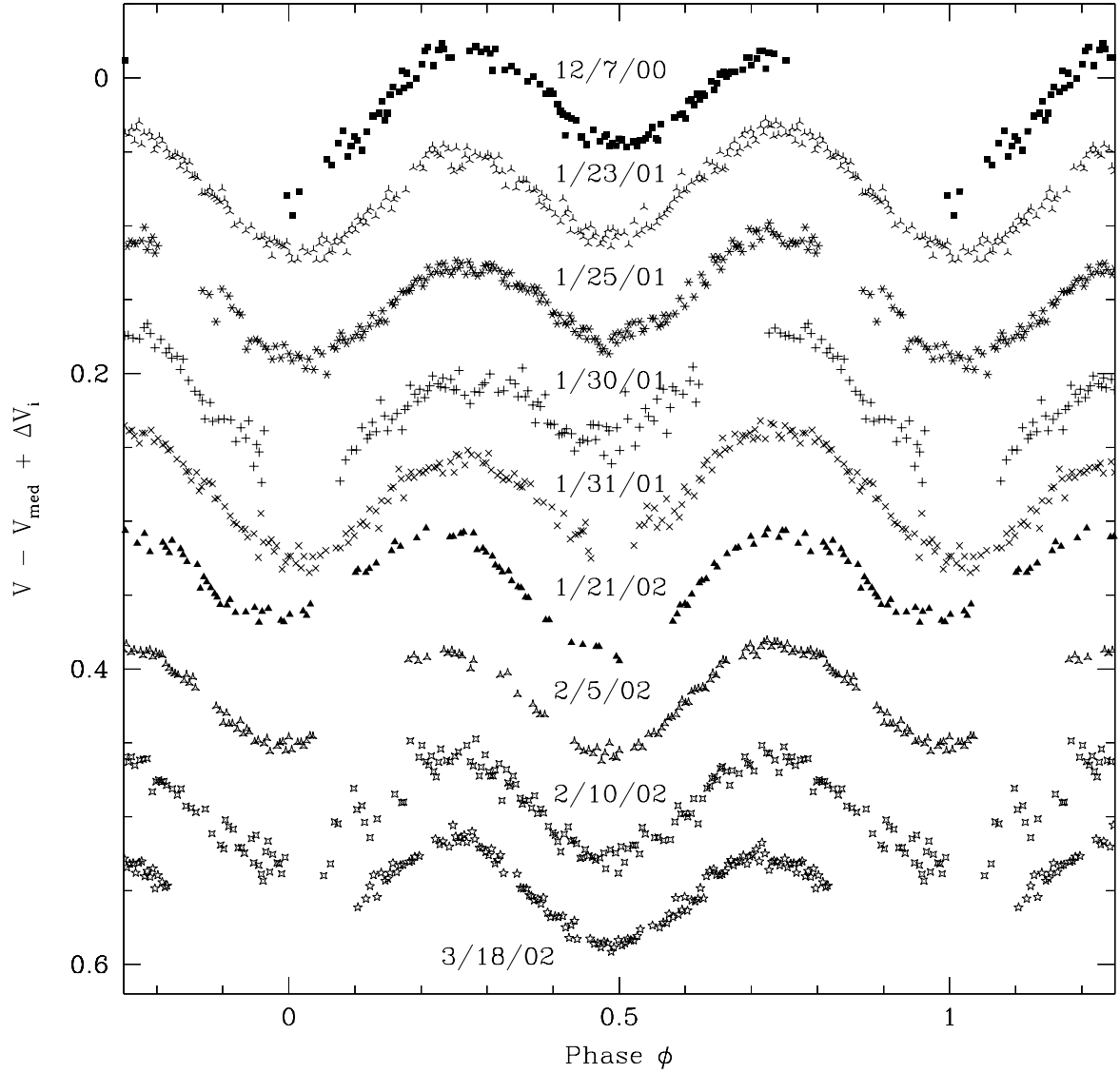}
\caption{The most complete single day $V$ light curves for the contact
binary S757. Successive light curves have been offset by 0.07
mag.\label{s757.2}}
\end{figure}

Figure ~\ref{s757.2} shows our most complete single-day light curves.
The observations from January 2001 in particular show significant
changes in the light curve from night to night.  The relative heights
of the maxima clearly changed between Dec. 7, 2000 and Jan. 30, 2001
with phase 0.25 slightly brighter than phase 0.75 in December, but
about 0.05 mag fainter in January. The magnitude level of the
secondary eclipse is the portion of the light curve that seems to
remain the most constant. The month timescale of these variations
leads us to believe that the cause is starspots. Spots have been
hypothesized on many of W UMa systems in connection with unequal
brightness of the two maxima (generally called the O'Connell effect;
O'Connell 1951). The case of S757 adds an additional wrinkle because
the maximum following the primary eclipse did change from being
brighter than the other maximum to being fainter (generally called a
variable O'Connell effect). This could indicate that either there are
significant spots present on {\it both} components or that hot {\it
and} cool spots appeared on one component of the binary. We tend to
believe the former explanation because the levels of the two maxima do
not appear to remain at constant brightness. Because we focused on
$V$-band observations, we do not have enough information to conduct a
more detailed analysis of the temperatures of hypothesized spots.  In
any case, the fact that we are unable to establish the true brightness
of the maximum of the light curve makes a derivation of reliable
system parameters nearly impossible.

In spite of the apparent spot activity, S757 was not detected as an
X-ray source in the cluster by Belloni, Verbunt, \& Mathieu
(1998). Both of the other W UMa systems of comparable brightness
(S1036 and S1282) were detected.  $L_{x} / L_{bol}$ increases with
increasing color for field stars, which implies that S 757 should be
brighter than S1282 in X-rays. The periods of the two systems are
very nearly the same, which means that differences in the rate of
rotation are not likely to account for differences in X-ray
activity. In addition, S757 falls in the region where the two X-ray
fields of Belloni et al. overlap, meaning that there was a longer
total integration on S757. There does not appear to be any
unidentified X-ray sources in the vicinity of S757, so we are unable
to explain the non-detection.

{\bf S1036 (EV Cnc):} S1036 is classified as a blue straggler
because accurate photometry shows it to be about 0.1 mag (in $B-V$) to
the blue of the cluster turnoff. Gilliland et al. (1991) determined
the period to be 10.59 h. vSVM redetermined the period to be $0.44144
\pm 0.00001$ d, and they note that this does not appear to be
consistent with the Gilliland et al. value, although they could not
judge how significant this was because Gilliland et al. did not quote
an error.  Using the vSVM data in combination with our own, we are
able to improve the determination of the period very slightly. Both
the L-K and L-S methods agree on slightly lower but different
periods. The L-S result appears to phase the data slightly better, and
the measured light curve parameters are presented in Table~\ref{wuma}.

\begin{figure}
\plotone{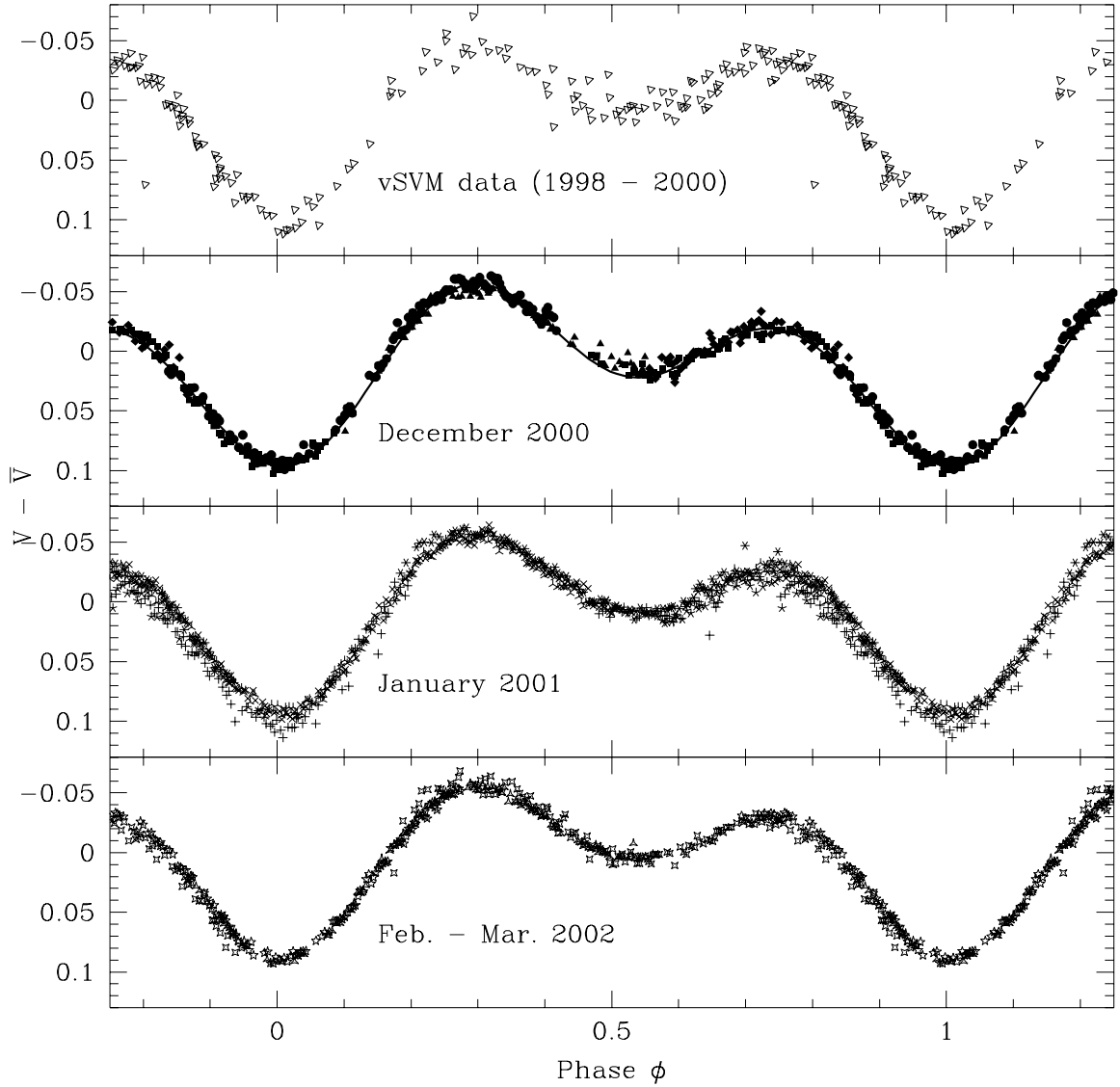}
\caption{Phased $V$ light curves for the contact binary S1036,
separated by month of observation. The top panel shows data from van
den Berg et al. (2002). Zero phase was chosen to be the photometric minimum.
Our best-fit light curve model is plotted with the December 2000 data.
\label{s1036}}
\end{figure}

The asymmetry in the light curve of S1036 makes it somewhat unusual:
for a contact binary with uniform surface temperature, the light-curve
maxima should have equal brightness. For S1036, the two maxima differ
in brightness by approximately 0.02 mag. The two light-curve minima
also differ in brightness by approximately 0.09 mag, as can seen in
Fig.~\ref{s1036}. For eclipsing binaries a difference in brightness of
the light curve minima normally implies a difference in temperature
for the two stars, something which is hard to understand if the stars
are in contact. In Table~\ref{wuma}, we include measurements of
eclipse depths $\Delta V_p = V_{p,min} - V_{max}$ and $\Delta V_s =
V_{s,min} - V_{max}$, where $p$ and $s$ refer to primary and secondary
minima, and $max$ refers to the global maximum of the light curve.

\begin{figure}
\plotone{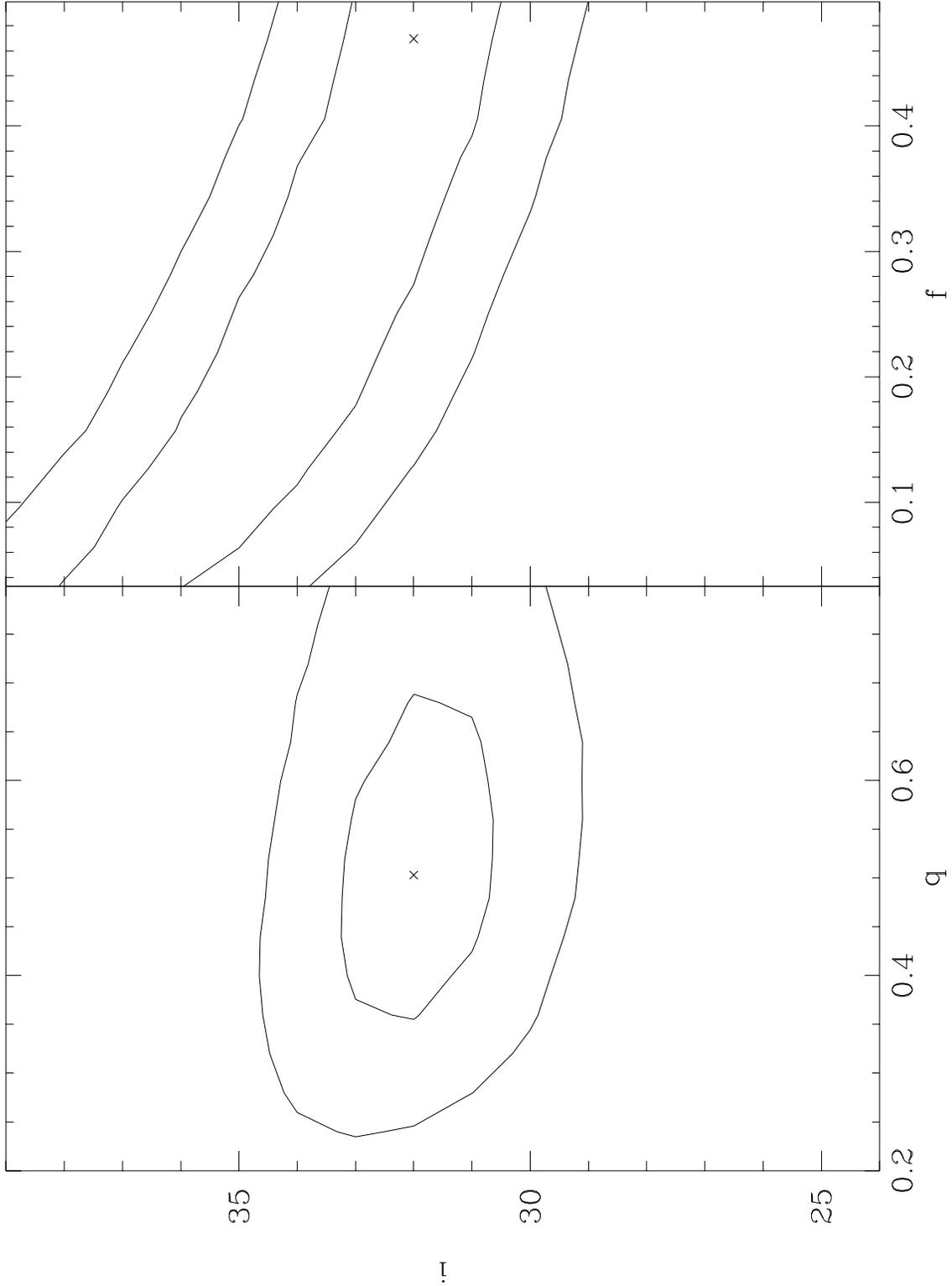}
\caption{$\chi^{2}$ contours for light curve fits to the December 2000
for S1036.  In both cases, the inner contour corresponds to
$\chi^{2}_{min} + 1.0$ and the outer contour corresponds to
$\chi^{2}_{min} + 4.0$, roughly 1 and $2 \sigma$ boundaries on $q$ and
$i$. $\times$ symbols indicate our baseline solution (described in the
text).\label{s1036chi}}
\end{figure}

We conducted limited modeling of S1036 using the program NIGHTFALL
(see \S \ref{lcanal} for details of the code). Two factors conspire to
prevent definitive determination of system parameters from the light
curve: low inclination and the possibility of spots. For warnings
about deriving system parameters from low-inclination systems, see
Rucinski (2001) and the discussion related to the system S1282 below.
The asymmetries in the light curves of W UMa variables have led many
researchers to model these systems with spots, usually with no more
than two spots. However, there are potentially many combinations of
spot parameters (spot latitude, longitude, size, and dimming factor)
that can model a single light curve. This is true of S1036
as well.  However, to check on the possibility that the determination
of a subset of the system parameters might be robustly determined from
these models, we first determined a set of system and spot parameters
that fits the system well ($\chi^{2} = 0.98$), and then proceeded to
compute grids of models varying two parameters at a time.  This
baseline solution had $q = 0.50$, $i=0.32$, $f = 0.47$, and two spots
(spot 1: facing us at phase 0.52, radius $30\degr$, dimming factor
0.72; spot 2: facing us at phase 0.23, radius $38\degr$, dimming
factor 0.80). The filling factor $f = (C_{1} - C) / (C_{1} - C_{2})$,
where C is the Jacobi potential of the common envelope, and $C_{1}$
and $C_{2}$ are the corresponding potentials of the $L1$ and $L2$
points We provide $\chi^{2}$ contour maps for two of these
experiments in Fig.~\ref{s1036chi}.

As can be seen in the figure, there is very little correlation between
$i$ and mass ratio $q$ (which is very poorly constrained). We checked
for correlations between inclination and spot parameters, and found
that they were minor. The primary degeneracy is between $i$ and the
filling factor $f$. The light curve clearly allows for solutions
covering the range of filling factors observed for most contact
binaries ($0 < f < 0.5$; Rucinski 1997). The reader should keep in
mind that the right panel of the figure was computed for a constant
set of spot parameters, and some adjustment of these could reduce the
$\chi^{2}$ value for a given combination of $i$ and $f$. Thus,
solutions with the component stars detached should very much be
considered to be viable. If the system is in contact, however, the
inclination can be constrained to be $30\degr < i < 38\degr$.

At least one spot appears to be necessary in models of the light curve
because the shallower minimum is displaced away from the expected
position at phase 0.5 and the two maxima still have different
brightness. vSVM saw no evidence for variation in color during the
orbit, although their quoted upper limit on the color variation was
0.05 in $B-V$. The limits on color variations placed by the vSVM data
do not distinguish between the possibilities that i) the surface
temperatures of the stars are equal and the differences in maximum and
minimum brightness are entirely due to cold spots, or ii) there is one
strong cold spot that explains the differences in maximum brightness
and a large temperature difference between the two stars that explains
the differences in minimum brightness. However, neither hypothesis
predicts color variations of more than about 0.01 mag, so much more
accurate color information is needed to constrain the models
strongly. As a result, we did not attempt to do a more systematic
study of the binary parameters using our photometry.  During our
observations, we did detect slight variations in the shape of the
light curve from month to month, although they are not as noticeable
as those seen for S 757. Given that the timescales for variations in
the light curves of the other W UMa variables discussed in this paper
are in the range of months or years, studies of the light curve of
S1036 over a longer time baseline could help determine whether
starspots are the primary factor in determining the light curve shape.

{\bf S1282 (AH Cnc):} With data from the studies of Whelan et
al. (1979) and Gilliland et al. (1991), there are currently good
datasets covering the entire light curve of this W UMa system for
three epochs separated by over a decade. Gilliland et al. noted that
both minima in the light curve appeared to have changed from curved
and continuously varying (possibly indicating partial eclipses) in the
Whelan et al. dataset to flatter (indicating total eclipses) in their
dataset. Our light curves closely resemble those of Gilliland et al.:
one flat-bottom eclipse covering approximately 0.1 in orbital phase,
and a slightly deeper and more curved primary eclipse.

\begin{figure}
\plotone{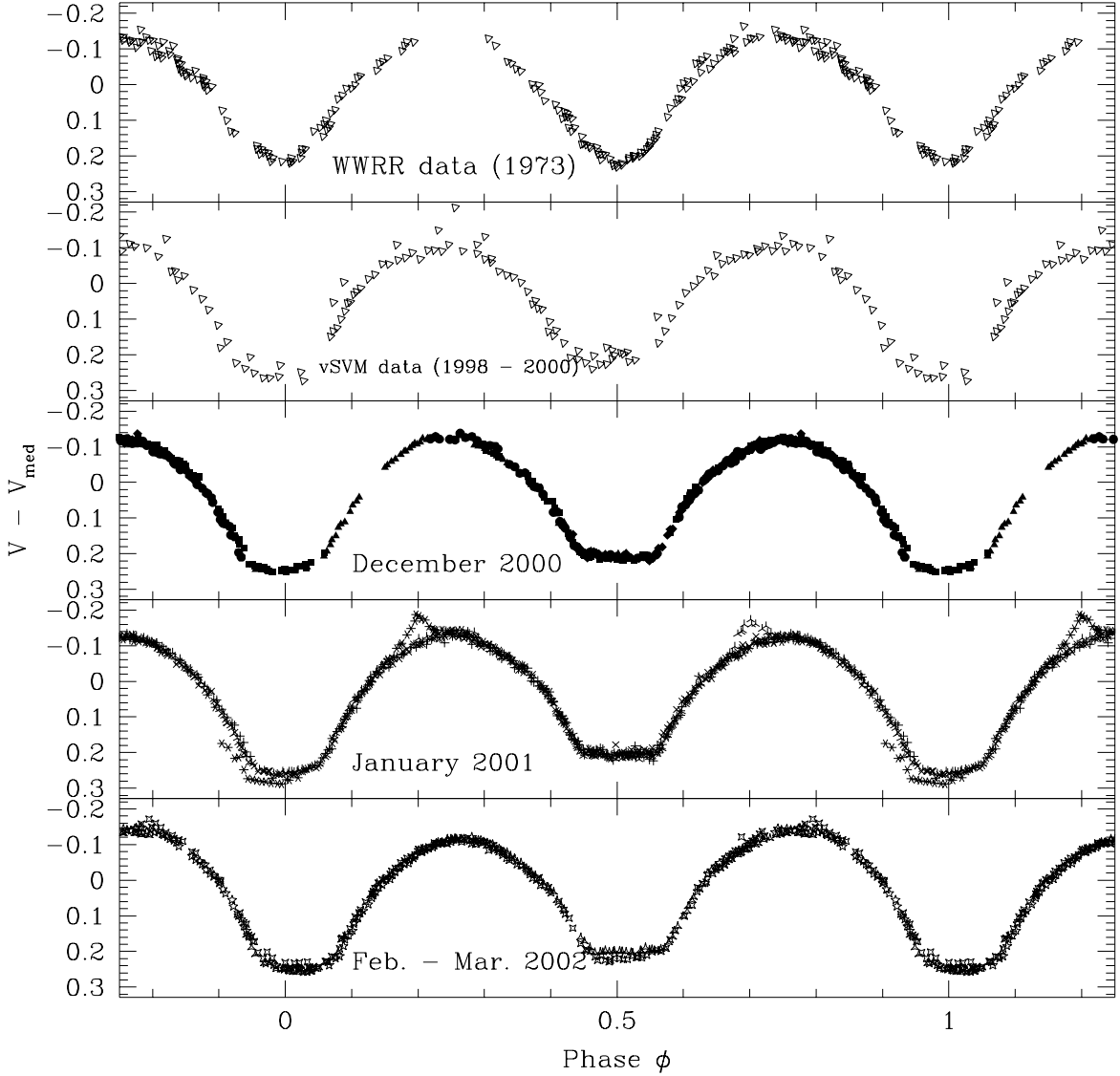}
\caption{Phased $V$ light curves for the contact binary S1282 (AH
Cnc) separated by month of observation. The top panel shows data from
Whelan et al. (1979), and next panel lower shows data from van den
Berg et al. (2002). Zero phase was chosen to be the total eclipse of the
least massive star.\label{s1282}}
\end{figure}

Figure~\ref{s1282} shows our observations along with those of vSVM and
\cite{whel79}. We have phased our observations using the ephemeris of
Kurochkin (1979), which includes a second-order time term, although we
have added a phase shift of 0.5 to bring the deeper minimum to phase
0. The deeper minimum in the Whelan et al. data is {\it not} the
deeper minimum in later data according to this ephemeris.  As can be
seen in the observations of Whelan et al., the minima were at
approximately the same brightness level in 1973. The Kurochkin
ephemeris does a good job of phasing the observations of vSVM with
ours, clearly illustrating the necessity of using the second-order
term.  We had to include an additional phase shift of 0.04 to our data
and those of vSVM in order to bring the primary minimum to phase 0,
which indicates that the ephemeris needs to be revised. [We should
note that the zero-point of the vSVM data appears to be different from
that of our data. Shifting their data according to the difference
between our median $V$ magnitude (13.44) and their average $V$
magnitude (13.54; SvMV) we find that the light curves differ by 0.05
mag. The difference is partly due to the difference between the
median and average for this light curve. For Figure~\ref{s1282}, we
simply added a magnitude correction to their data to make
comparison of the light curves easier.]

Based on the appearance of the eclipses in our data, the system is
clearly an A-type W UMa binary (the deeper eclipse being a transit of
the larger star by the smaller star, as seen from the curvature of the
light curve near phase 0) rather than W-type as identified by Whelan
et al. Given the position of S1282 near the turnoff of M67 in the
color-magnitude diagram, it might have been predicted that it should
be an A-type system since they are generally believed to have evolved
components (Mochnacki 1981). The relationship between type and
evolutionary state may be misleading, however, since S1282 appears to
have effectively changed type in less than a decade (between the
observations of Whelan et al. and those of Gilliland et al.). One
should also look at the data for the systems S757 (discussed above)
and III-79 (discussed below), which show that one eclipse minimum can
go from being the fainter of the two to the brighter in a matter of
months or years.  Because this is much too short to correspond to any
reasonable evolutionary timescale (nuclear, thermal, or even
dynamical) for the stars themselves, we believe that this shows that
the evolutionary status of a W UMa variable cannot be determined
without an extended study of the light curve and its potential for
variation. For S1282, we do have additional information that is not
always available for a W UMa system: the binary is found at the
turnoff of the cluster, and so the primary star is likely to have
depleted much of its core hydrogen supply.

Because S1282 was close to our primary target (S1082) in the cluster,
we have photometry from most nights of observations, and on some
nights we were able to cover almost an entire orbital period.
Previous modeling based on the Whelan et al. data (Maceroni et al
1984) indicated that the system significantly overfills the Roche
lobes of the two stars. However, the presence and duration of the
total eclipse invalidates the low inclination ($i = 62\fdg9$) that
Maceroni et al.  derived. Given that the Whelan et al. observations of
light curve minima occurred over a relatively short period of 2.5
months, it is possible that there was a short-term change in the light
curve shape. We have observed shorter timescale changes to the light
curve shape (for example, the deeper primary eclipse observed on
January 25, 2001), and the variability seen by vSVM in their secondary
eclipse observations during two runs seems to support this.  However,
the amplitude of the light curve from Whelan et al. seems to be
somewhat smaller than the amplitude in our data, which would tend to rule out
cool spots as the cause.

While Whelan et al. did present radial velocity measurements in order
to measure the spectroscopic mass ratio, they were not able to
constrain the velocity semi-amplitude of the secondary star
precisely. The large differences between their light curve and later
ones might also indicate that their radial velocities may have been
systematically affected by spots (to give just one possibility). As a
result, we will not take their spectroscopic mass ratio to be a
primary constraint on light curve models. The radial velocity
measurements were good enough to show, however, that the less massive
star is totally eclipsed as it passes behind the more massive star, as
would be expected if the more massive star is also larger in size.

In most cases, W UMa light curves by themselves provide very little
information about the system parameters, and attempts to derive
parameters using $\chi^{2}$ fits are extremely dangerous (see Rucinski
2001 for a discussion). However, when a contact system shows total
eclipses, the duration of totality strongly constrains the combination
of mass ratio and inclination, and is insensitive to the degree of
contact (Mochnacki \& Doughty 1972). Because S1282 clearly has total
eclipses, a fairly stable light curve during the period of our
observations, and has not been modelled previously, light-curve models
can provide valuable constraints.

We modeled the light curve of S 1282 using NIGHTFALL (see \S
~\ref{lcanal} for details). We focus on data from December 2000 and
January 2001 because nearly the entire light curve was observed in
relatively short periods of time. We decided not to attempt to model
data taken in 2002 because the maximum following primary eclipse was
found to be significantly fainter ($0.02 - 0.04$ mag) than the maximum
following secondary eclipse, probably indicating that spots were a
significant influence.

\begin{figure}
\plotone{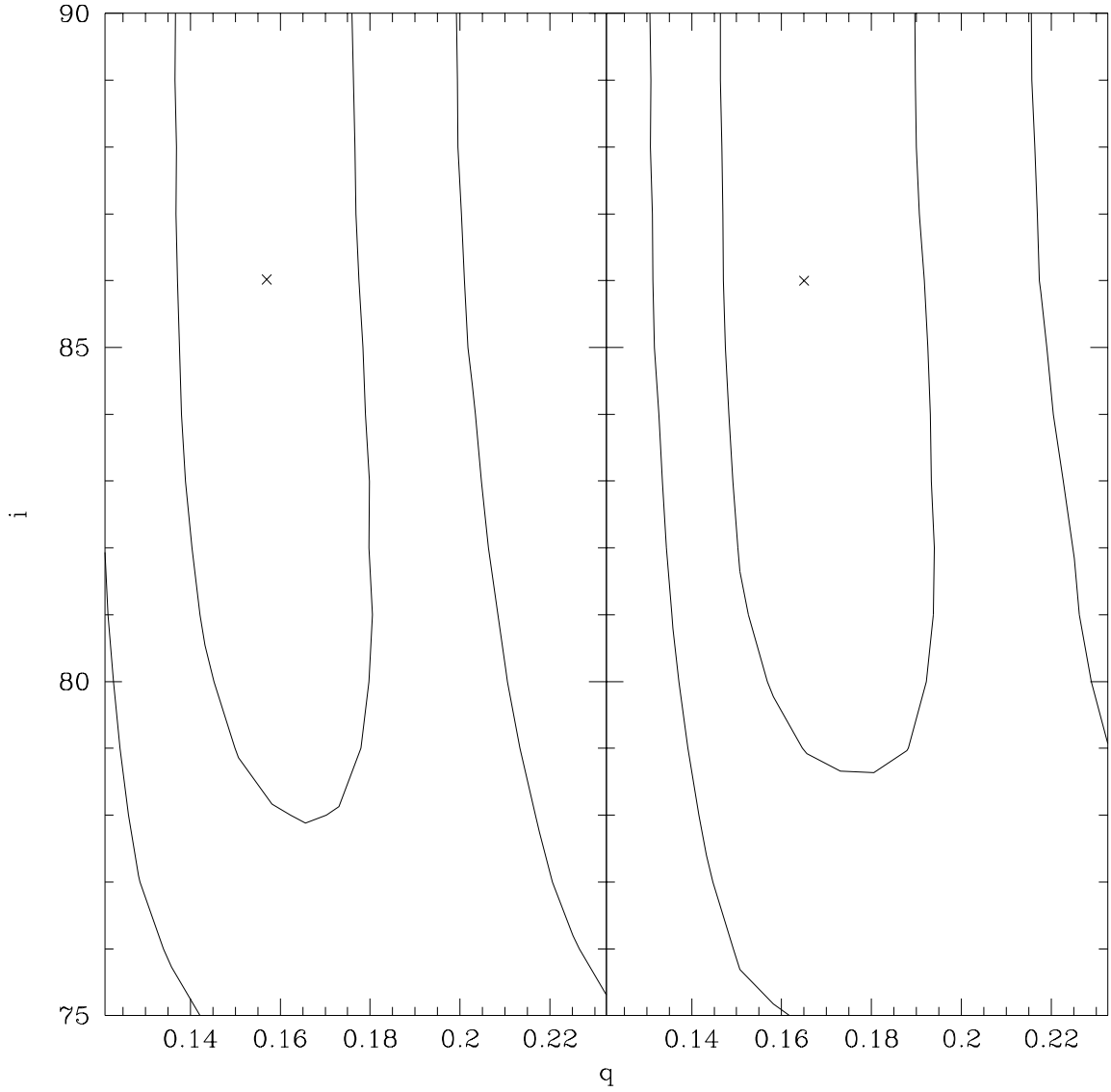}
\caption{$\chi^{2}$ contours for light curve fits to the December 2000
({\it left panel}) and January 2001 ({\it right panel}) data for S1282.  In
both cases, the inner contour corresponds to $\chi^{2}_{min} + 1.0$
and the outer contour corresponds to $\chi^{2}_{min} + 4.0$, roughly 1
and $2 \sigma$ boundaries on $q$ and $i$. $\times$ symbols indicate
the best fit solutions.\label{chis}}
\end{figure}

The large width of both eclipses requires an inclination near $90
\degr$ and a small mass ratio.  Our best-fit
values for the December 2000 and January 2001 data separately are
presented in Table~\ref{tfits}. In order to estimate the possible
errors, we ran grids of models varying $q$ and $i$ to determine
changes in $\chi^{2}$. The resulting contour maps are shown in
Fig.~\ref{chis}. The contours correspond to levels 1.0 and 4.0 above
the minimum $\chi^{2}$ fit for that dataset, which should roughly
delineate the 1 and $2 \sigma$ confidence regions $q$ and $i$ taken
individiually. As seen in the contour maps, there is little
correlation between $q$ and $i$. The best fits contours for the two
datasets agree well, although there is a small shift in $q$ between
the results for the two datasets.  Using the information from the
$\chi^2$ maps and the systematic shifts between the best fit models
for the two datasets, we quote best fit values of $q =
0.16^{+0.03}_{-0.02}$ and $i = 86 \degr ^{+4}_{-8}$. The variations in
the best-fit parameters due to month-to-month changes in the light
curve appear to be comparable in importance to the random errors in
the fitted value of $q$.

\begin{figure}
\plotone{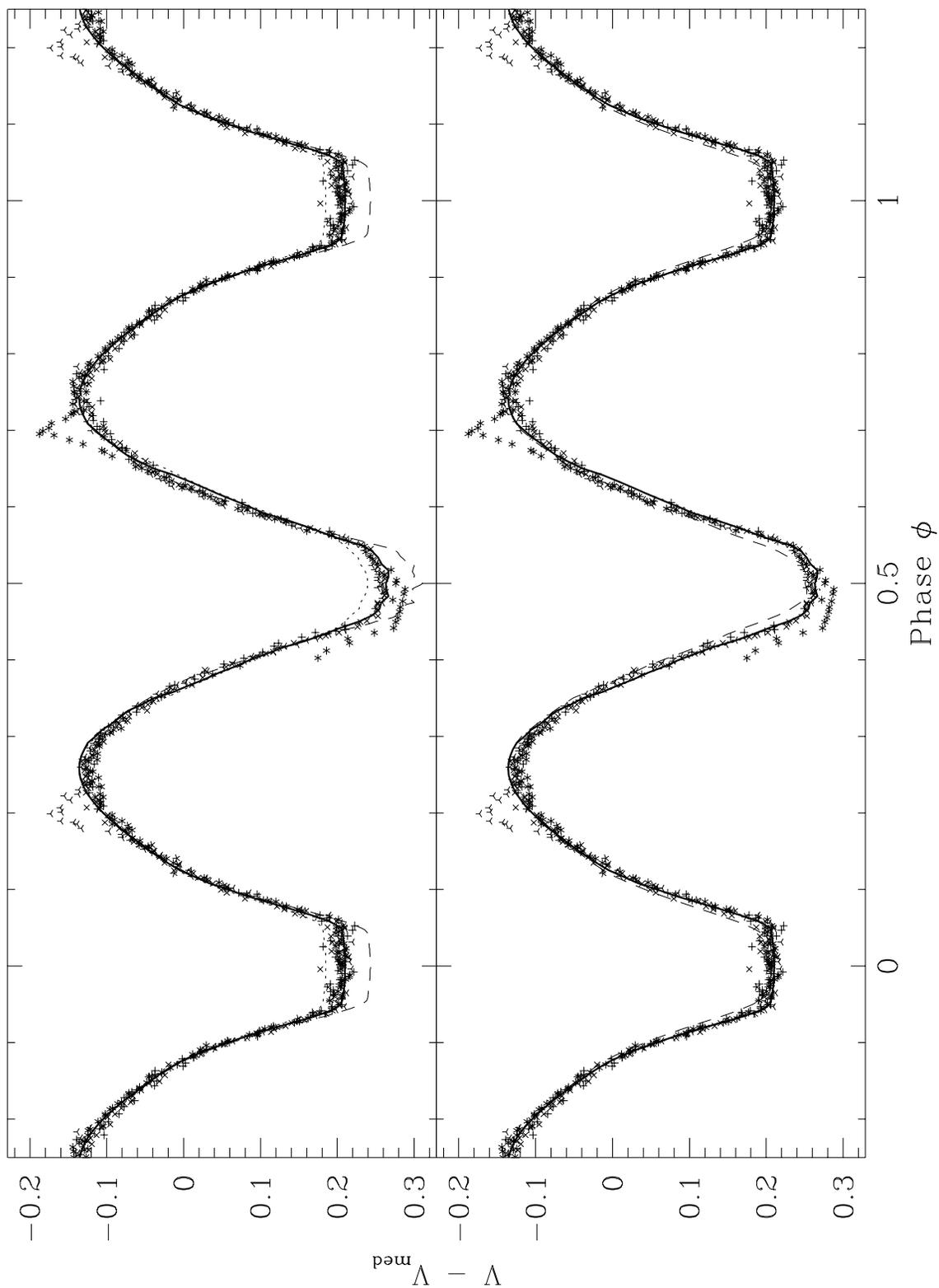}
\caption{Trial light curve fits to the January 2001 data for S1282.
In both panels, the heavy solid line is our best fit model ($q=0.165$,
$i=86\degr$).  Models with constant $i$ are shown in the upper panel:
$q = 0.145$ (dotted line) and $q=0.195$ (dashed line). Models with
constant $q$ are shown in the lower panel: $i=90\degr$ (dotted line)
and $i=78\degr$ (dashed line). \label{fits}}
\end{figure}

In Fig.~\ref{fits}, we show a selection of theoretical models against
the observational data. The best fit model is almost lost among the
observational points, with the poorest fit near $\phi = 0.25$. The
upper panel of the figure shows two models with the best-fit
inclination but with mass ratios $1 \sigma$ away from the best
model. The mass ratio is primarily constrained by the eclipse depths.
The lower panel shows light curves with the best-fit mass ratio, but
with different inclinations. The $i = 90\degr$ curve is almost
indistinguishable from the best fit model, but the $i = 78\degr$ curve
has eclipses of noticeably shorter duration.

Although the derived mass ratio for the system falls near the low end
of the distribution for W UMa systems, W UMa systems tend to be found
preferentially at low $q$ values (Rucinski 2001), so this fact is not
unusual. The filling factor for the system is constrained by
the shape of the light curve before and after primary eclipse,
and although it is correlated with the inclination (in the sense that
lower inclinations require larger filling factors), there is negligible 
degeneracy between mass ratio and filling factor. For both of our
best-fit models we find it necessary to use a filling factor ($f
\approx 0.7$) that is probably larger than the majority of observed
systems (Rucinski 1997). A lower limit on the filling factor is $f > 0.4$
based on $\chi^2$ values measured for models with $i = 90\degr$.

The results of the light curve analysis contradict the earlier
spectroscopic analysis of \citet{whel79}. Although their derived
radial velocities were very uncertain ($K_{1} \approx 100 \pm 15$
\kms, 138 \kms $< K_{2} <$ 240 \kms), the implied mass ratios are
inconsistent with our photometric value. In addition, if it is assumed
that the maximum $K_{2}$ value is approximately correct (giving a
spectroscopic $q$ value closest to our photometric value), the derived
total system mass is substantially less that that predicted for a
turnoff star in M67 from stellar evolution models ($\sim 1.25 \msun$)
in spite of the system's position near the turnoff in the
color-magnitude diagram. We are forced to conclude that the radial
velocity data for this system are not currently of high enough quality
to derive trustworthy masses.  The importance of a good radial
velocity curve for this system should be emphasized, since it would
provide a valuable constraint on the cluster turnoff mass, and would
thereby help more accurately age-date M67.

\begin{figure}
\plotone{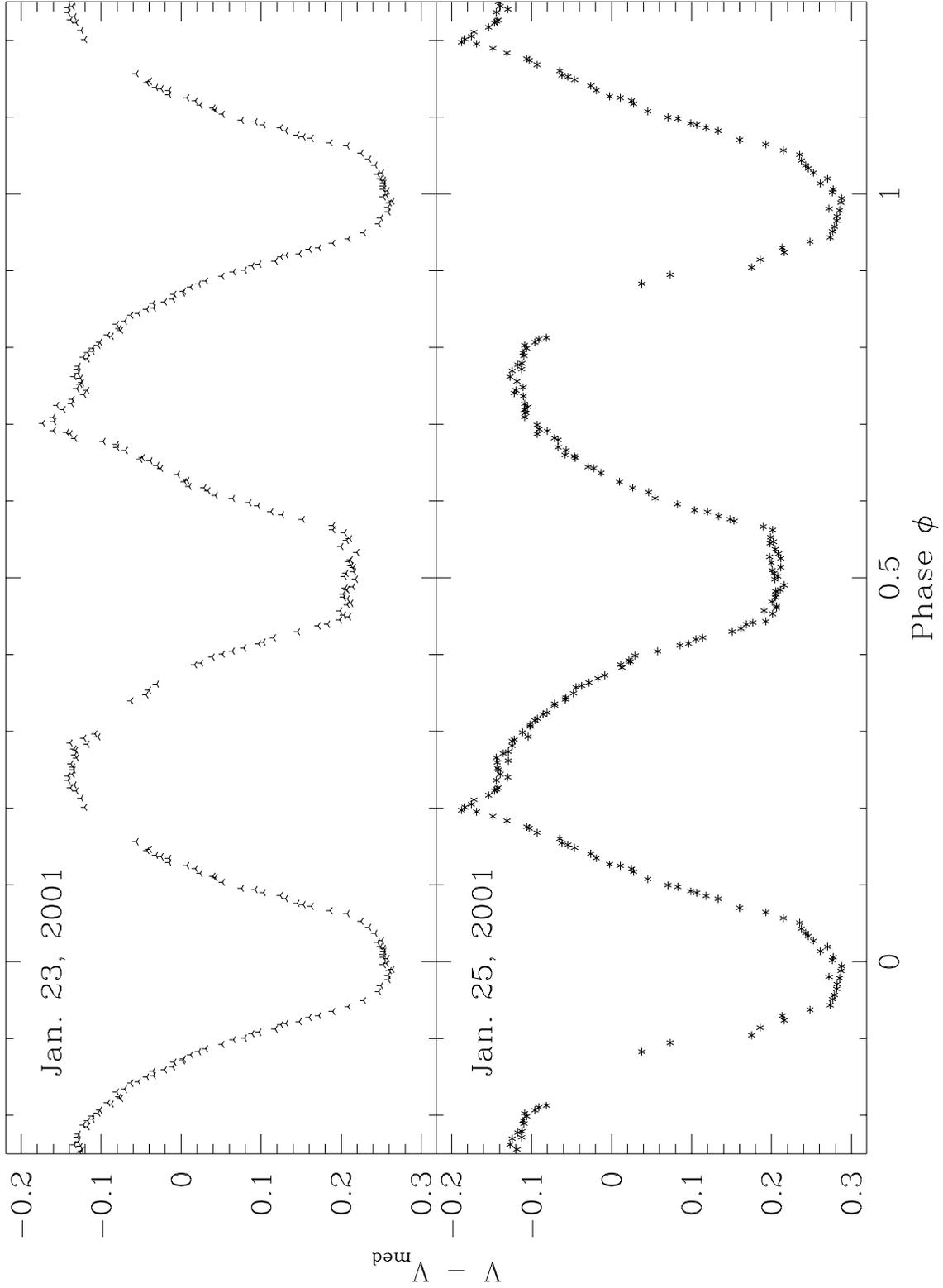}
\caption{Disturbed $V$ light curves for the contact binary S1282 (AH Cnc)
in January 2001.\label{weird}}
\end{figure}

Our extensive observations allowed us to discover some unusual
transient features in the light curves (Fig.~\ref{weird}). On January
23/24 and 25/26, 2001, we observed short ($\sim
30$ min) brightness increases ($\sim 0.08$ mag for both, compared to
eclipse depths of 0.33 and 0.39 mag). Unusual features of these two
brightenings were: they were observed at almost identical phase
positions shortly before maximum, the following maximum did not show
the feature, and on the two days the brightness increase was seen on
{\it different} maxima.  In addition, the primary minimum preceding
the brightness increase on Jan. 25/26 was significantly deeper (by $0.04
- 0.05$ mag) than any other minimum observed. These features
disappeared within two days.

These transient features bear some resemblance to variations seen in the light
curve of the nearby W UMa variable 44i Boo (e.g. Duerbeck 1978), which
appears to have active periods of a few years duration interspersed
with quieter periods with undisturbed light curves. A similar cycle
(probably related to the magnetic field) might help to explain the
difference between the light curves of \citet{whel79} and other
observers.

The timing of the features shortly before maximum seems to argue
against stellar flares. The short duration of the features argues
against hot spots, which should persist for longer times. One
hypothesis is based on magnetically-confined gas (prominences) on one
of the stars on the portion of the surface facing away from the other
star. If the gas has sufficient optical depth, it could potentially
increase the effective surface area of the binary when a large slab is
aligned edge-on, but could cause either modest or negligible dimming
of the system when viewed face-on through the thin dimension of the
slab. The increased depth of the first half of the primary eclipse on
January 25 indicates that the feature was probably on the side of the
secondary star facing almost directly away from the primary. Because
our light curve was taken entirely in $V$, we have no direct
information on temperature.  (We wish to thank M. Blake for the
suggestion that initiated this line of thought.)

\begin{figure}
\plotone{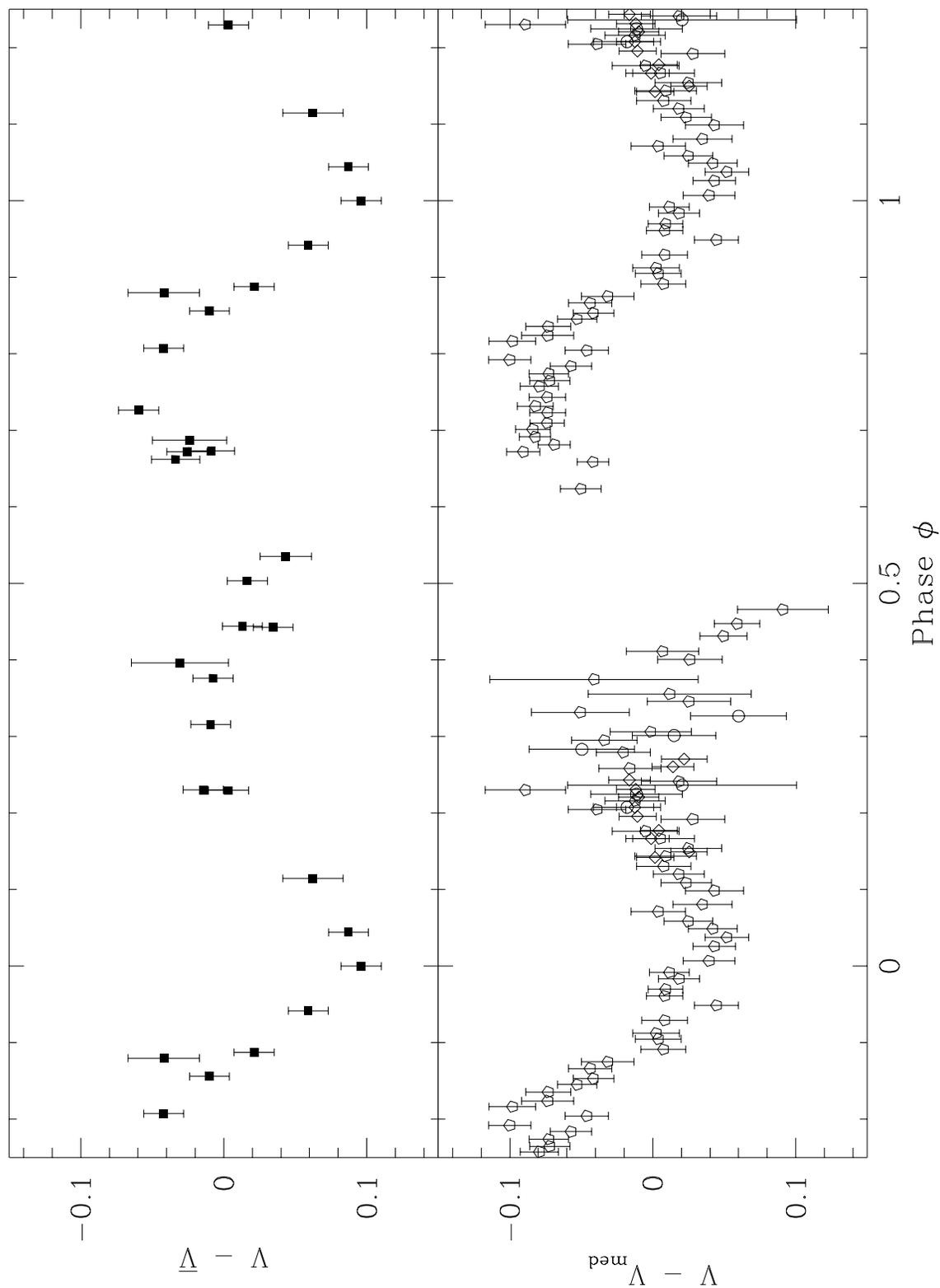}
\caption{Phased $V$ light curves for the contact binary III-79 (ET
Cnc). The upper panel shows data from Stassun et al. (2002), and the
lower panel shows our measurements from March 2001. Both datasets have
been phased to the same linear ephemeris ($P = 0.270505$ d). Zero
phase was chosen to be the photometric minimum of the Stassun et
al. dataset.\label{etcnc}}
\end{figure}

{\bf III-79 (ET Cnc):} This system was discovered by Gilliland et
al. (1991) at $7\arcmin$ from the cluster center, and was also
observed by SvMV. This binary is considerably fainter than the other
three contact systems known. This system was in the field of our March
2001 observations, with the best data coming from night 13. We
averaged observations to increase the S/N ratio in the data.  Our data
are presented in Fig.~\ref{etcnc} along with those of SvMV from
January 1998, phased to the same linear ephemeris using a period ($P =
0.270505$ d) that is consistent with the value quoted by SvMV to
within the errors. Our measured median magnitude ($V = 15.90$) agrees
very well with the average magnitude of SvMv (15.89), so no shifting
in magnitude was done.

Although individual data points have considerably larger errors than
those for the other W UMa systems discussed, a comparison of the light
curves indicates a noticeable change in the light curve shape. While
the brighter of the two light curve maxima follows the brighter
minimum in the SvMV and Gilliland et al. datasets, the brighter
maximum appears to follow the fainter minimum in our data. Although we
do not have data completely covering the minimum plotted at phase
$\phi = 0.5$, several points with good errors indicate that it has
dimmed noticeably. Independent of this, the $\phi = 0.5$ minimum does
not reach the same depth as the $\phi = 0$ minimum in the light curves
of Gilliland et al. and SvMV. So the indication is that this system
shows fairly extreme variations in the shape of the light
curve. Without more information about the nature of this variation, it
would be reckless to try to model this system.

\subsection{Blue Stragglers}

During the course of the observations we observed most of the
cluster's blue stragglers during the majority of at least one night in
order to look for variability on various timescales. The results
of this search are given in Table~\ref{straggler}. Our list of
stragglers comes primarily from the list of Ahumada \& Lapasset
(1995), although stars have been excluded if the accurate photometry
of \citet{fan96} indicated that the star was close to the cluster
turnoff. Stars S2223 and S2226 were added to the list based on the
photometry of \citet{fan96} and proper motion membership
\citep{sanders77,girard89}.

We have monitored a number of stragglers for the first time in this
study, but we do not find any of these to be noticeably variable. We
provide the values for our variability indicators $\sigma_{V}$ and
$I_{WS}$ (discussed in \S \ref{vari}) in columns 4 and 5 respectively.
Typically nonvariable stars in the magnitude range of the stragglers
had $\sigma_{V} < 0.015$ and $I_{WS} < 1.5$. We discuss most of the
variability candidates in \S \ref{otherv} below. We note that both
indicators can give erroneously high scores if observations were taken
during poor seeing conditions and a contaminating star was present
nearby, as was the case for the straggler S975. We have provided
information on the nights the stars were observed in order to lay the
groundwork for constraining variations on longer timescales. We do,
however, rule out the possibility that any of the other blue
stragglers observed are W UMa variables of amplitude more than about
$0.01 - 0.02$ mag.

\subsubsection{$\delta$ Scuti Stars}

The blue stragglers in the CMD cover the region where the instability
strip occurs. We have roughly translated the instability strip of
Pamyatnykh (2000) into our observational CMD, as shown in
Fig~\ref{fig1}. We will discuss our oscillation mode analyses of the
two known blue straggler $\delta$ Scuti stars S1280 and S1284 in a
separate paper in preparation. However, we find several other
stragglers with photometry that places them inside the instability
strip: S1263, S1267, S968, S1066 and S1434 to the blue of the known
pulsators, and S752, S2226, S1082 and S975 to the red. We did not
observe S1434 or S2226. S752, S975, and S1267 are known to be long
period variables ($P = 1003$ d, 1221 d, and 846 d, respectively;
Latham \& Milone 1996), while S 1082 contains a close binary that may
be part of a triple \citep{vdb01,sand02}. Therefore, these systems are
less likely to show photometric variations due to pulsation (although
it is possible that components of the long period variables will still
fall in the instability strip: in particular, the brightest component
of S 1082 shows some evidence of being a $\delta$ Scuti star;
Sandquist et al. 2003). As a result, S 968, S 1066, and S 1263 are the
best candidates to search for pulsation.  \citet{gill92} observed S
1263 extensively, and also found no sign of variation. We do not find
convincing evidence of pulsation in any of the three stars, although
this may be due to low pulsation amplitudes or higher frequency
(overtone) oscillations. We discuss all three stars in \S \ref{otherv}
with regards to longer period variations.

\subsubsection{Possible Variables}\label{otherv}

As discussed earlier, to tightly constrain the photometric variations
present in some of the bluest blue stragglers, we were forced to
subtract off a systematic error (probably related to color response of
the filter/CCD combination) that appeared as correlated slow
low-amplitude variations among several stars. Once this correction was
made several of the stars no longer showed any signs of significant
variation (S 977, S 1066, S 1267). The remaining three (S 752, S 968,
S 1263) still showed trends, and will be discussed below. None of
these stars tripped the variability condition $I_{WS} \ga 1.5$,
although in the case of S 752 at least there is fairly clear evidence
of variability in one interval of time. Before presenting the results, we note
that we looked for correlations of the magnitude residuals with seeing
and airmass for each star below, and found no evidence that such
effects were responsible.

\begin{figure}
\plotone{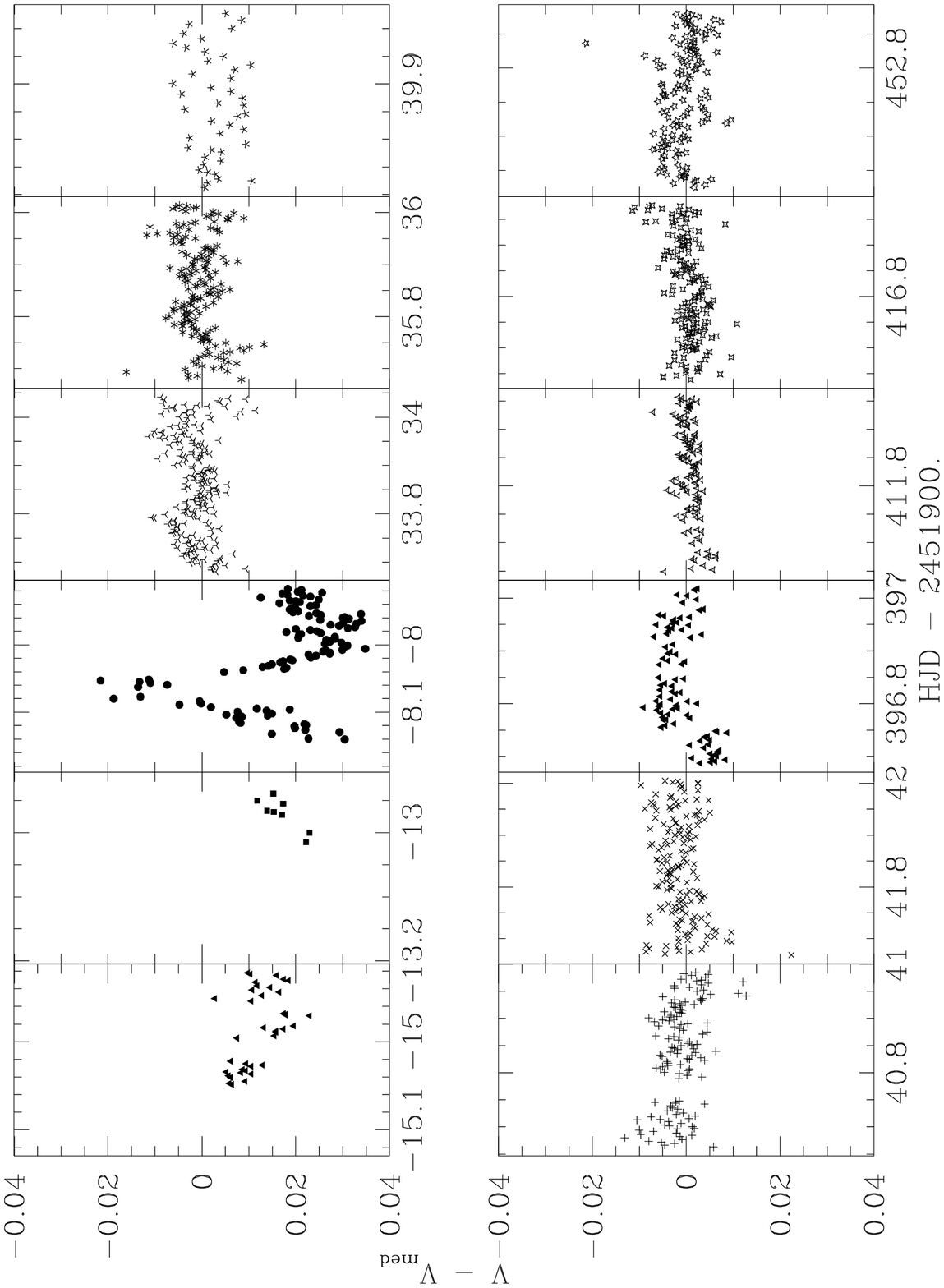}
\caption{$V$-band time-series photometry for the blue straggler
S752. Tick marks on the time axis are spaced by 0.05 d.\label{s752}}
\end{figure}

{\bf S 752:} As noted earlier, S 752 is known to be a binary system
($P = 1003$ d; $e = 0.32 \pm 0.12$; Latham \& Milone 1996) containing
an Am star. This star was outside the field observed by
\citet{gill91}, and although it was observed by SvVM, it was not
reported as variable. Simoda (1991) observed the star and found it to
be non-variable. We present our observations in Fig.~\ref{s752}.
While the vast majority of our measurements show no indication of
variability, our December 2000 observations give the indication of a
decrease in brightness of 0.03 mag over the course of a week,
punctuated with something resembling a flare (0.05 mag increase in
brightness lasting over 3 hours). However, see Gilliland et al. 1991
for a discussion of false flaring.

\begin{figure}
\plotone{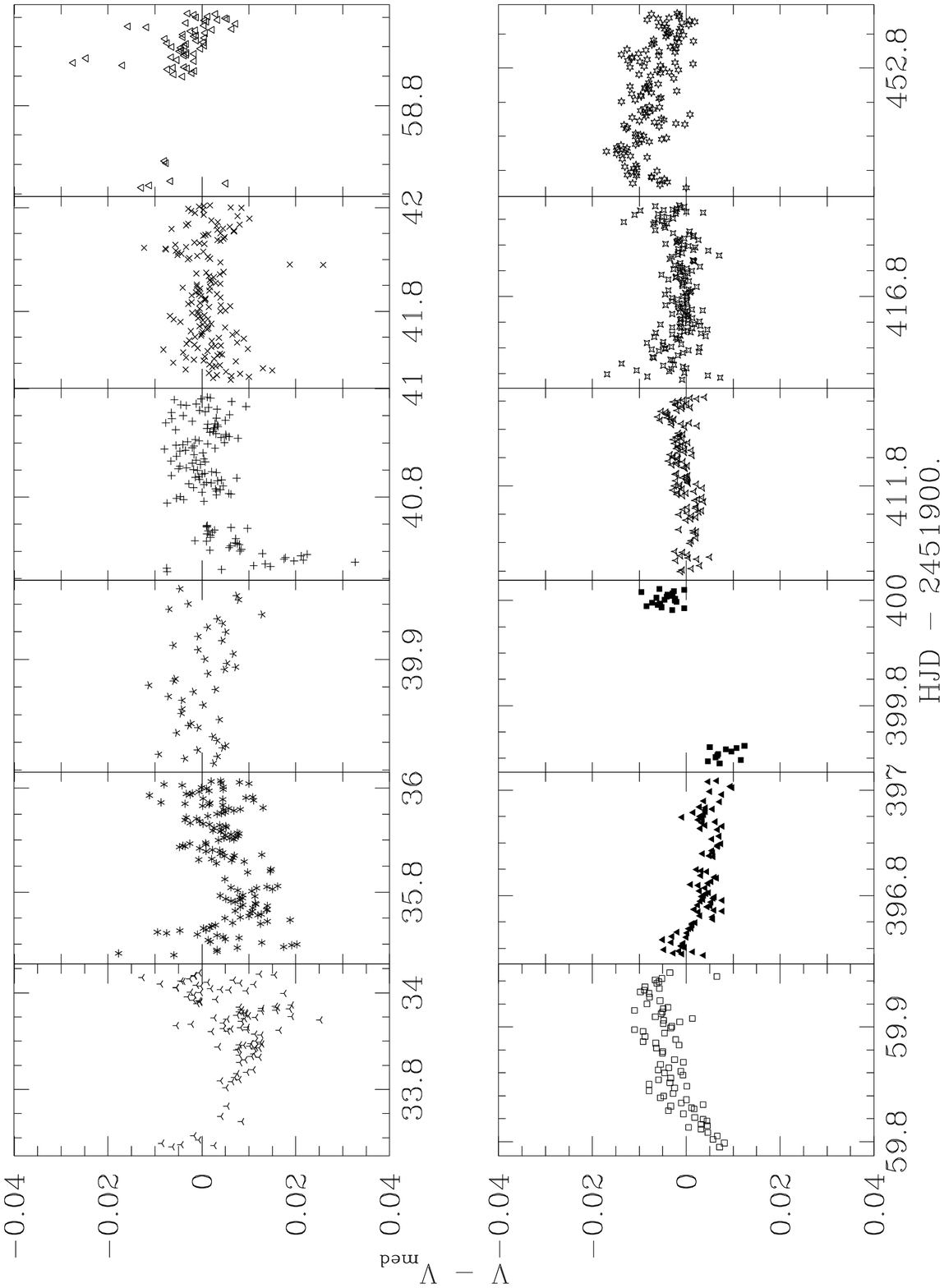}
\caption{$V$-band time-series photometry for the blue straggler
S968. Tick marks on the time axis are spaced by 0.05 d.\label{s968}}
\end{figure}

{\bf S 968:} This blue straggler was not in the field observed by
\citet{gill91}, but was observed by SvMV although they did not report
it as being variable. Our observations are shown in
Fig.~\ref{s968}. For most of the nights this star was observed, it was
fairly consistent with constant brightness. On other nights there
appeared to be noticeable trends during a night (most notably February
18/19, 2001). This behavior could explain the non-detection by
SvMV.

\begin{figure}
\plotone{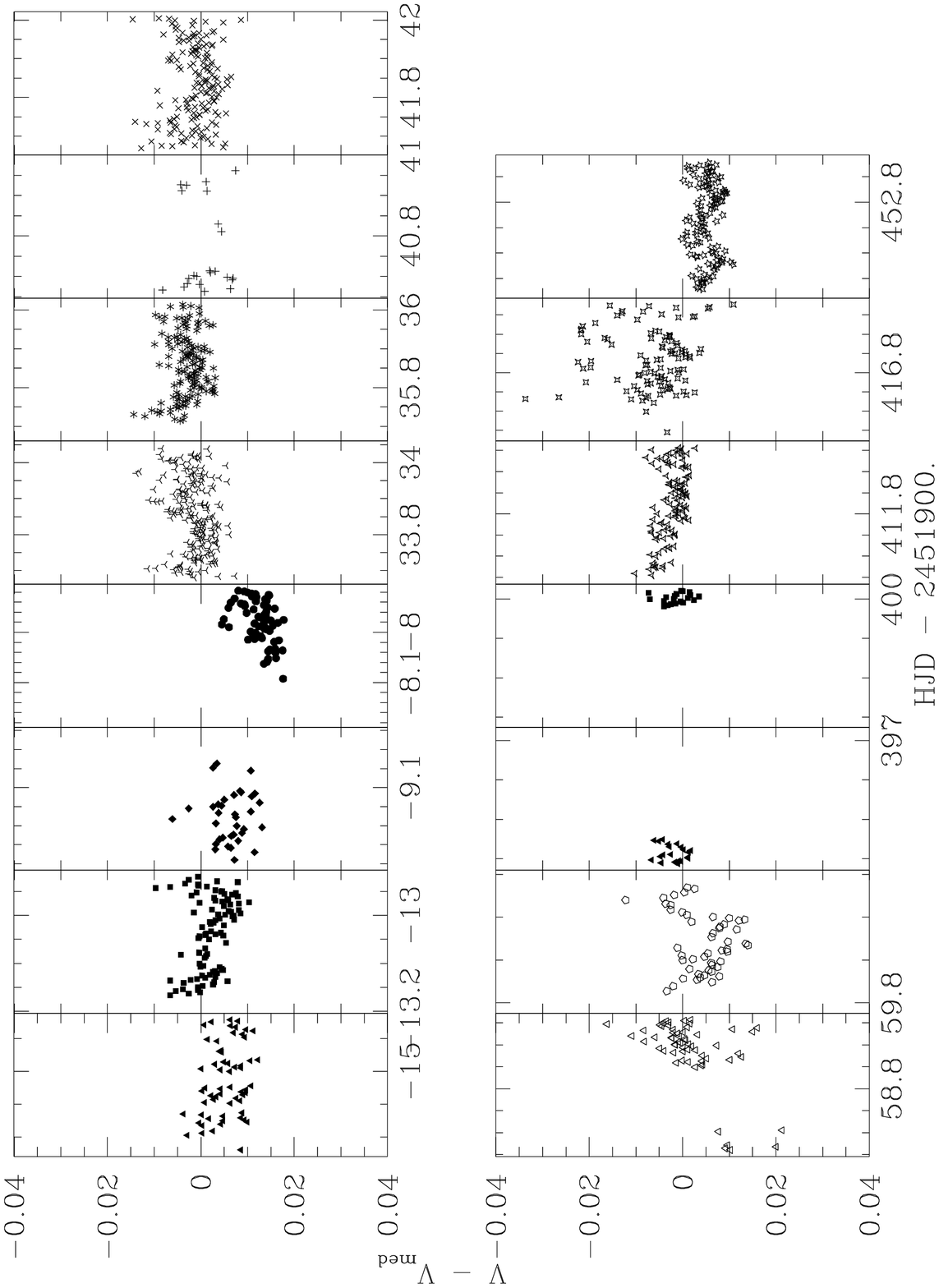}
\caption{$V$-band time-series photometry for the blue straggler
S1263. Tick marks on the time axis are spaced by 0.05 d.\label{s1263}}
\end{figure}

{\bf S 1263:} SvMV indicate that their observations of this star have
significant scatter of up to 0.03 mag, and Kim et al. (1996) find
large dispersion in their measurements. There does appear to be a long
term drift in the brightness of the star (of approximately 0.02 mag)
as seen in Fig.~\ref{s1263}, even after removing the effects of
correlated residuals. Some of our nights of observations were
compromised by very poor seeing and contamination by MMJ 5951
\citep{montgomery93}, which is about $9\arcsec$ away. Based on trends
seen in the photometric residuals in measurements during the poorest
seeing, we eliminated measurements with seeing greater than $4\farcs8$
FWHM. We attempted to look for periodic signals in the data, although
we did not find convincing evidence of any (the best indications were
for a period of 17.5 d). This system is not known to be part of a
binary, so it should be monitored further to determine the manner of
its variation.

\subsection{Long Period Variables}

{\bf S 1040:} This system was determined by Mathieu, Latham, \&
Griffin (1990) to be a single-lined spectroscopic binary having a
circular orbit and a period $P = 42.8271 \pm 0.0022$ d. The position
of S 1040 to the blue side of the red giant branch in the
color-magnitude diagram makes it a ``red straggler''. Using
ultraviolet imaging and spectra, \citet{land97} showed that the
secondary companion to the red giant primary is likely to be a helium
white dwarf. \citet{land97} also laid out a hypothetical history for the
system in which a main sequence binary began mass transfer at a
period of approximately 2 d when the more massive star had a helium
core mass of $0.16 \msun$. Mass transfer continued until the envelope
of the initially more massive star was depleted, creating a system
composed of a helium white dwarf and a blue straggler. Once the blue
straggler evolved to the red giant branch, a system like S 1040 would
be created.

\begin{figure}
\plotone{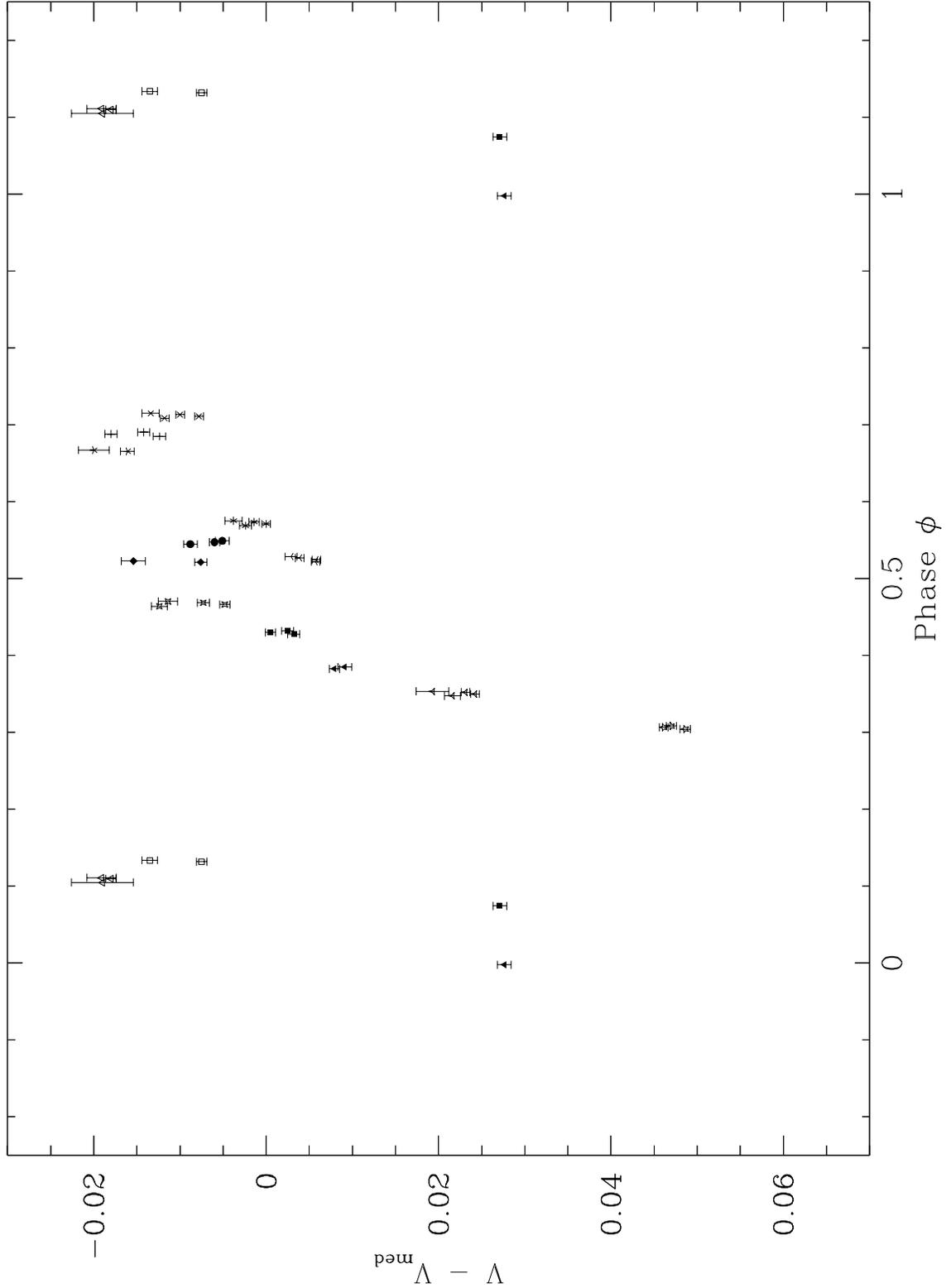}
\caption{The $V$ light curve for the ``red straggler'' binary S1040.
Observations have been averaged in bins 0.1 d in duration. The data
has been phased to the spectroscopic ephemeris of Mathieu, Latham, \&
Griffin (1990). Phase $\phi = 0$ corresponds to maximum positive
radial velocity of the brighter star in the system.\label{s1040}}
\end{figure}

In previous variability studies, \citep{gill91} found evidence for low
amplitude (0.012 mag) variation with a period of 7.97 d, while vSVM
found no evidence of variation. We plot our data (averaged in 0.1 d
bins to improve the errors) in Fig.~\ref{s1040}, and tabulate it in
Table \ref{s1040tab}. We have observed several nights for which the
system brightness decreased by as much as 0.06 mag compared to the
peak level. When the data is phased to the ephemeris of Mathieu et
al. (1990), there is an indication of a systematic decrease in
brightness at a phase corresponding to the passage of the white dwarf
in front of the giant. Because our phase coverage is incomplete, we
are unable to determine the exact depth of the feature. However, a
feature of this depth and shape cannot be explained as an eclipse of
the red giant by the white dwarf. Observations on two nights do not
match up well with the light curve formed from the rest of the
observations. We probed a variety of periods from 1 to 70 d, but did
not find satisfactory light curves from any of the best trial periods
from the L-K or L-S methods. Additional observations should be made to
better determine the shape of the light curve, and the nature of the
variation.

{\bf S 1063:} This system is a binary with orbital period of $18.396
\pm 0.005$ d and eccentricity $0.206 \pm 0.014$ (Latham et al. 1992;
Mathieu et al. 2002), although an orbit has not yet been published for
the system. S 1063 is known to be unusual in its X-ray emission
\citep{bell98} and position in the color-magnitude diagram (fainter
than the subgiant branch). There is good evidence of proper motion
membership though \citep[$> 90\%$ probability]{sanders77,girard89}.

This system fell just outside the Gilliland et al. field, but was
observed by vSVM in their examination of X-ray sources. vSVM
identified the system as as a non-periodic variable based on an
examination of periods up to the length of their longest interval of
continuous observations (18 d). Their Figure 5 makes it clear that at
the very least there is a great deal of variability in the light
curve, and that the variation is not obviously related to the orbital
characteristics of the system.

\begin{figure}
\plotone{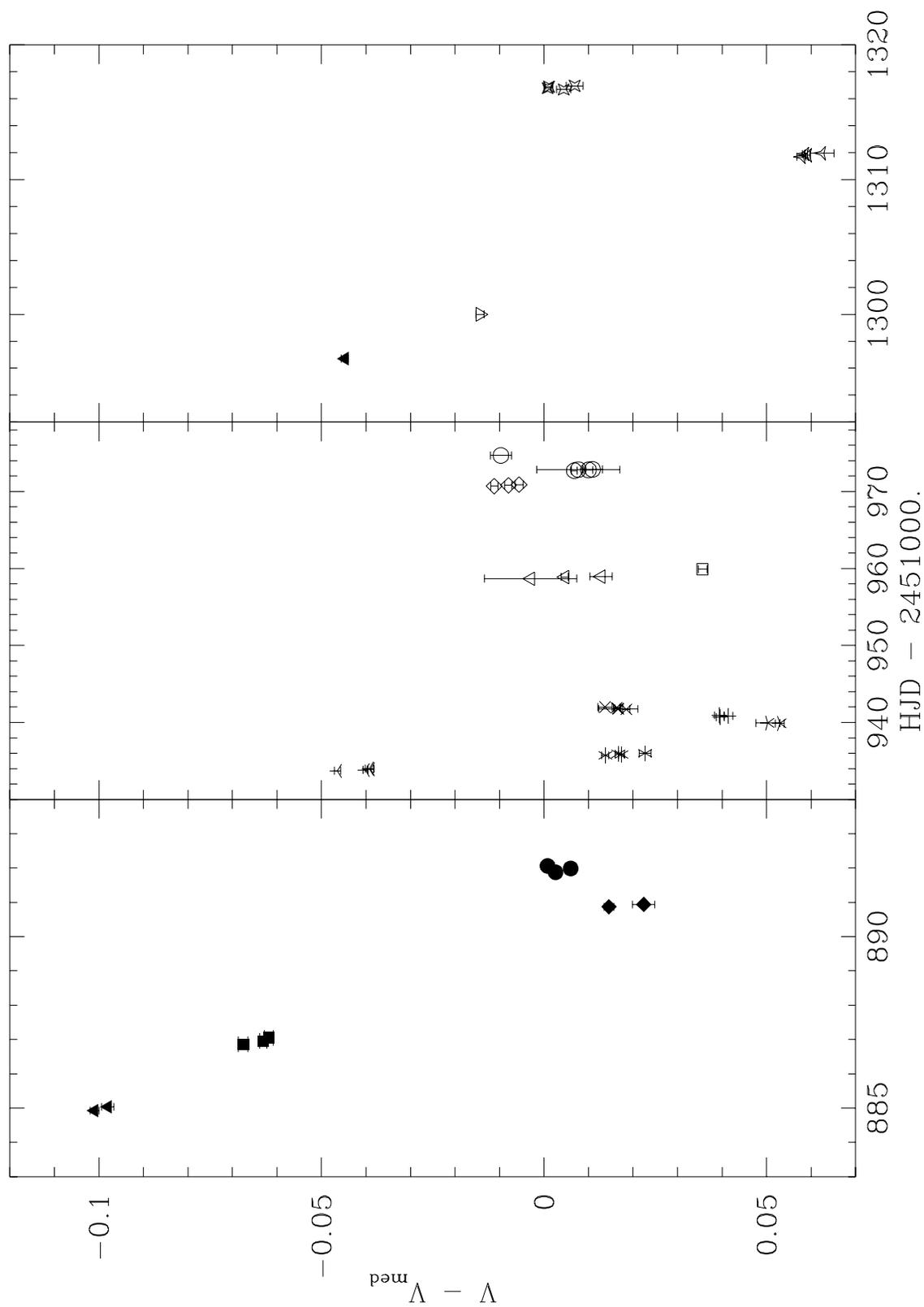}
\caption{$V$-band time-series photometry for the sub-subgiant branch
binary S1063.  Observations have been averaged in bins 0.1 d in
duration. Note that the time axis in the middle panel is more
compressed than those on either side.\label{s1063}}
\end{figure}

We averaged observations from a given night in 0.1 d windows to
improve the error in our measurements, and they are presented in
Table~\ref{s1063tab}. In our data, the system shows slight photometric
trends during the course of a night, but systematic offsets in the
star's photometry are clearly detected from night to night. Using an
L-S periodogram, we do not find any evidence of variation on the
orbital timescale, but there is slight evidence of variation with a
period of approximately 23 days. This period seems to bring the maxima
and minima of our data into approximate alignment. As shown in
Fig.~\ref{s1063}, a comparison of data from different observing runs
indicates that the brightness level of the minimum of the light curve
probably varies a few centimag from cycle to cycle. During the course
of our observations the light curve appeared to maintain a roughly
similar shape, but one that appears to differ from the portions
observed by vSVM. Although the system does not show periodicity on the
orbital period, the similarity of the timescale may indicate that the
variability is related via tides and/or magnetic activity. Further
progress on this system will require observations over much
longer periods and a more detailed examination of the relationship
with the orbit.

[We have three nights of observations of another sub-subgiant branch
system (S 1113), but our phase coverage for this system is fairly poor,
so we do not discuss it.]

\section{Conclusions}

We have presented $V$ observations for the open cluster M67, and have
discussed the light curves for the known W UMa contact binaries and
the monitoring of the majority of the blue stragglers for variability.
We find that all of the known W UMa binaries show light curve
variations that occur on timescales of days to months. Two systems
(S757 and III-79) show large changes in the shapes of their light
curves. The relative brightnesses of both the two maxima and the two
minima in the light curve of S757 have been observed to change on
timescales of less than a month. The faint system III-79 shows a
substantial change in the shape of the light curve between the January
1998 of SvMV and our observations in March 2001. The other two systems
(S1036 and S1282) show smaller variations. S1282 changed between a
W-type and an A-type configuration between the 1973 observations of
Whelan et al. (1979) and the 1988 observations of Gilliland et
al. (1991).  The existence of two systems which appear to change
between these subtypes indicates that the classification is perhaps
not a robust indicator of the evolutionary state of the stellar
components. The blue straggler system S 1036 shows smaller light curve
variations, but a stronger and more stable O'Connell effect.

We have verified that S 1282 is a highly-inclined totally-eclipsing
system with $q = 0.16^{+0.03}_{-0.02}$ and $i = 86\degr^{+4}_{-8}$. Because
this system falls right at the cluster turnoff, we strongly encourage
further spectroscopic work to provide a constraint on the cluster
turnoff mass, and thus on the cluster age. This system has also shown
unusual short disturbances in its light curve that may relate to
magnetic activity, and which should be followed up.

Among the blue stragglers, in addition to the two known $\delta$ Scuti
pulsating variables, we find possible evidence of longer period
variations in the stars S752, S968, and S1263. While these stars
did not satisfy our criteria for a definite claim of variability, we
see trends in the photometry and apparent changes in the mean
brightness level that should be investigated.

Finally, we present a series of observations of two long-period binary
systems. For the poorly understood sub-subgiant branch system S 1063,
there are indications of quasi-periodicity on timescales similar to
the orbital period, as well as variations in the mean brightness level
of the light curve. For the giant-white dwarf system S 1040, we find
evidence of a drop in brightness at phases corresponding to the
passage of the white dwarf in front of the giant using the ephemeris
derived by Mathieu et al. (1990). Although this drop in brightness
cannot be due to an eclipse of the giant by the white dwarf itself,
there may be associated material within the system that could account
for the variability. This information can probably be used to contrain
the inclination of the system.

\acknowledgments

We would especially like to thank M. van den Berg and K. Stassun for
the loan of data from their photometry papers.  E.L.S. would like to
thank M. Blake, S. Rucinski, W. Welsh, R. Taam, and P. Etzel for
helpful conversations during the course of this work. We would also
like to thank the director of Mount Laguna Observatory (P. Etzel) for
generous allocations of telescope time. This work has been partly
supported by NSF grant AST-0098696 to E.L.S.

\newpage

\clearpage
 
\clearpage
 
\clearpage
 
\clearpage
 
\clearpage
 
\clearpage
 
\clearpage
 
\clearpage
 
\clearpage
 
\clearpage
 
\clearpage
 
\clearpage
 
\clearpage
 
\clearpage
 
\clearpage
 
\clearpage
 
\begin{deluxetable}{cccc}
\hspace*{-0.2in}
\tablecolumns{4}
\tablewidth{0pc}
\tablecaption{Observing Log for $V$ Photometry at Mt. Laguna}
\tablehead{\colhead{\#} & \colhead{Date} & 
\colhead{mJD Start\tablenotemark{a}} & \colhead{$N_{obs}$}}
\startdata
1 & Dec. 5/6, 2000 & 1884.878 & 60\\
2 & Dec. 7/8, 2000 & 1886.806 & 111\\
3 & Dec. 11/12, 2000 & 1890.817 & 57\\
4 & Dec. 12/13, 2000 & 1891.820 & 116\\
5 & Jan. 23/24, 2001 & 1933.675 & 206\\
6 & Jan. 25/26, 2001 & 1935.673 & 176\\
7 & Jan. 29/30, 2001 & 1939.676 & 44\\
8 & Jan. 30/31, 2001 & 1940.657 & 130\\
9 & Jan. 31/Feb. 1, 2001 & 1941.663 & 161\\
10 & Feb. 17/18, 2001 & 1958.655 & 63\\
11 & Feb. 18/19, 2001 & 1959.790 & 75\\
12 & Mar. 1/2, 2001 & 1970.619 & 205\\
13 & Mar. 3/4, 2001 & 1972.631 & 205\\
14 & Mar. 5/6, 2001 & 1974.721 &  32\\
15 & Jan. 21/22, 2002 & 2296.687 & 89\\
16 & Jan. 24/25, 2002 & 2299.690 & 124\\
17 & Feb. 5/6, 2002 & 2311.649 & 119\\
18 & Feb. 10/11, 2002 & 2316.638 & 167\\
19 & Mar. 18/19, 2002 & 2352.625 & 131\\
\enddata
\label{obs}
\tablenotetext{a}{mJD = HJD - 2450000}
\end{deluxetable}

\begin{deluxetable}{lcccc}
\tablewidth{0pt}
\scriptsize
\tablecaption{Properties of M67 W UMa Systems}
\tablehead{
\colhead{ID}&
\colhead{S757} &
\colhead{S1036} &
\colhead{S1282} &
\colhead{III-79} \\
\colhead{}&
\colhead{}&
\colhead{EV Cnc}&
\colhead{AH Cnc}&
\colhead{ET Cnc}
}
\startdata
$T_{0}$ & 2451800.129 & 2450500.047 & \tablenotemark{e} & 2450000.036 \\
$P$ (d) & $0.35967 \pm 0.00002$ & $0.441437 \pm 0.000003$ & 0.360452
\tablenotemark{ce} & 0.2704 \tablenotemark{d} \\
$\Delta V_{p}$ & 0.08 & 0.13 & 0.39 & $\ga 0.17$\\
$\Delta V_{s}$ & 0.06 & 0.06 & 0.33 & 0.11 \\
$\overline{V}$ \tablenotemark{a} & 13.51 & 12.81 & 13.44 & 15.90 \\
$(B-V)$ & 0.61 \tablenotemark{c} & 0.50 \tablenotemark{b} & 0.52 \tablenotemark{b} & 1.11 \tablenotemark{d} \\
$i$ & & $\sim35\degr$ & $86\degr^{+4}_{-8}$ & \\
$q$ & & & $0.16^{+0.03}_{-0.02}$ & \\
\enddata
\label{wuma}
\tablenotetext{a}{Average $V$ magnitudes from observations in this
study.}  
\tablenotetext{b}{Stassun et al. (2002)}
\tablenotetext{c}{van den Berg et al. (2002)}
\tablenotetext{d}{Gilliland et al. (1991)} 
\tablenotetext{e}{Period known to vary with time; see ephemeris of
Kurochkin (1979)}
\end{deluxetable}

\begin{deluxetable}{lcc}
\tablewidth{0pt} 
\tablecaption{Best Fits to the Light Curve of S1282} 
\tablehead{\colhead{Quantity}& \colhead{Dec. 2000} & \colhead{Jan. 2001}} 
\startdata 
$i$ & $86\fdg0$ & $86\fdg0$ \\ 
$q$ & 0.157 & 0.165 \\ 
$f$ & 0.73 & 0.62\\
$\chi^{2}_{min}$ & 0.41 & 1.31 \\ 
\enddata
\label{tfits}
\end{deluxetable}

\begin{deluxetable}{ccrcrlrl}
\tablewidth{0pt} 
\scriptsize 
\tablecaption{Blue Straggler Observations} 
\tablehead{ \colhead{ID\tablenotemark{a}} & \colhead{$V$} & \colhead{$B-V$} &
\colhead{$\sigma_{V}$} & \colhead{$I_{WS}$} 
& \colhead{Nights Observed\tablenotemark{b}} & \colhead{$N_{obs}$} 
& \colhead{Results}} 
\startdata 
145 & & & & & none & & \\ 
277 & & & & & none & & \\ 
751 & 12.70 & & 0.008 & 0.53 & $5-9$ & 777 & no variation\\ 
752\tablenotemark{c} & 11.32 & 0.30 & 0.018 & 0.75 & $4-9,15,17-19$ & 1418 & Am star; 
possible flare detected\\ 
792 & 11.99 & 0.60 & 0.007 & 1.01 & $5-9,12,13,17-19$ & 1610 & no variation \\ 
968\tablenotemark{c} & 11.25 & 0.13 & 0.010 & 0.74 & $5-9,11,15,17-19$ & 1261 & Am star; possible low amplitude, \\
 & & & & & & & timescale $\sim$ days \\
975 & 11.05 & 0.43 & 0.077 & 7.26 & $1-9,11,15,17-19$ & 1659 &
photometry affected by faint companion \\
977\tablenotemark{c} & 10.02 & $-0.07$ & 0.006 & 0.69 & $5-9,18,19$ & 504 & no variation\\
984 & 12.26 & 0.57 & 0.014 & 1.64 & $1-9,11,15,17-19$ & 1732 & no variation \\
997 & 12.13 & 0.46 & 0.006 & 0.92 & $1-9,11,15,17-19$ & 1655 & no variation \\
1005 & 12.68 & 0.52 & 0.010 & 1.57 & $1-9,11,15,17-19$ & 1654 & no variation \\
1031 & 13.29 & 0.46 & 0.007 & 0.32 & $1-9,11,15,17-19$ & 1644 & no variation \\
1036 & 12.81 & 0.49 & 0.046 & 90.59 & $1-9,11,15,17-19$ & 1651 & W UMa variable (EV Cnc)\\
1066\tablenotemark{c} & 10.95 & 0.11 & 0.003 & 0.24 & $1-9,11-13,17-19$ & 2061 & no variation \\ 
1072 & 11.31 & 0.61 & 0.006 & 0.97 & $1-9,11-13,17-19$ & 2073 & no variation\\
1082 & 11.20 & 0.42 & 0.024 & 33.59 & $1-9,11-13,17-19$ & 2097 & RS CVn variable (ES Cnc)\\
1165 & & & & & none & & \\ 
1183\tablenotemark{d} & 12.66 & & 0.011 & 0.07 & 15 & 66 & no variation\\
1195 & 12.29 & & 0.003 & 0.40 & 15 & 78 & no variation\\ 
1263\tablenotemark{c} & 11.06 & 0.19 & 0.007 & 0.98 & $1-9,11,17-19$ & 1275 & possible low amplitude,\\
 & & & & & & & timescale $\ga$ 10 days\\
1267\tablenotemark{c} & 10.90 & & 0.005 & 0.45 & $1-4,11$ & 484 & no variation\\ 
1273 & 12.25 & 0.57 & 0.006 & 0.63 & $1-9,11-13,17-19$ & 1451 & no variation\\
1280 & 12.23 & 0.26 & 0.009 & 2.55 & $1-9,11-13,17-19$ & 1685 & $\delta$ Scu variable (EW Cnc) \\
1282 & 13.44 & 0.52 & 0.133 & 108.08 & $1-14,17-19$ & 1864 & W UMa variable (AH Cnc)\\
1284 & 10.93 & 0.22 & 0.012 & 9.41 & $1-9,11-13,17-19$ & 2015 & $\delta$ Scu variable (EX Cnc)\\ 
1434 & 10.70 & 0.11 & & & none & & \\ 
1440 & & & & & none & & \\
1947 & & & & & none & & \\ 
2204 & 12.89 & 0.45 & 0.013 & 1.03 & $1-9,11,15,17-19$ & 1717 & no variation \\
2223 & 13.30 & 0.50 & 0.012 & 0.92 & $1-9,11-13,17-19$ & 2084 & no variation \\ 
2226 & & & & & none & & \\ 
\enddata
\label{straggler}
\tablenotetext{a}{ID from Sanders (1977).}
\tablenotetext{b}{Night IDs given in Table~\ref{obs}}  
\tablenotetext{c}{Photometry corrected for color-related errors at the 0.02 mag level.}  
\tablenotetext{d}{May be a normal turnoff star.}  
\end{deluxetable}

\begin{deluxetable}{cccr}
\tablewidth{0pt}
\scriptsize
\tablecaption{Averaged Photometry for S1040}
\tablehead{\colhead{mJD\tablenotemark{a}}&
\colhead{$\bar{V}$}& \colhead{$\sigma_{\bar{V}}$} & \colhead{$N_{obs}$}}
\startdata
1884.9294 & 11.5231 &  0.0006 &   33\\
1885.0391 & 11.5243 &  0.0008 &   27\\
1886.8586 & 11.5184 &  0.0007 &   36\\
1886.9615 & 11.5157 &  0.0006 &   43\\
1887.0468 & 11.5177 &  0.0007 &   32\\
1890.8688 & 11.5076 &  0.0007 &   42\\
1890.9386 & 11.4998 &  0.0014 &   14\\
1891.8734 & 11.5064 &  0.0008 &   37\\
1891.9872 & 11.5092 &  0.0006 &   51\\
1892.0607 & 11.5101 &  0.0008 &   28\\
1930.7983 & 11.5395 &  0.0058 &    1\\
1933.7305 & 11.5209 &  0.0005 &   55\\
1933.8308 & 11.5210 &  0.0005 &   63\\
1933.9312 & 11.5191 &  0.0005 &   57\\
1934.0118 & 11.5181 &  0.0007 &   31\\
1935.7288 & 11.5128 &  0.0007 &   45\\
1935.8314 & 11.5152 &  0.0005 &   60\\
1935.9333 & 11.5138 &  0.0006 &   54\\
1936.0000 & 11.5114 &  0.0010 &   17\\
1939.8739 & 11.4991 &  0.0008 &   36\\
1939.9386 & 11.4952 &  0.0018 &    8\\
1940.7053 & 11.5028 &  0.0007 &   37\\
1940.8386 & 11.4972 &  0.0007 &   46\\
1940.9371 & 11.5010 &  0.0007 &   47\\
1941.7175 & 11.5034 &  0.0005 &   49\\
1941.8186 & 11.5074 &  0.0005 &   48\\
1941.9199 & 11.5052 &  0.0005 &   50\\
1941.9883 & 11.5018 &  0.0010 &   14\\
1958.6849 & 11.4962 &  0.0036 &    7\\
1958.8989 & 11.4972 &  0.0006 &   50\\
1958.9531 & 11.4961 &  0.0017 &    5\\
1959.8451 & 11.5077 &  0.0006 &   47\\
1959.9218 & 11.5017 &  0.0009 &   28\\
2296.7175 & 11.5428 &  0.0008 &   21\\
2300.0020 & 11.5423 &  0.0008 &   20\\
2311.6980 & 11.5368 &  0.0009 &   23\\
2311.7991 & 11.5394 &  0.0005 &   44\\
2311.8994 & 11.5383 &  0.0005 &   46\\
2311.9536 & 11.5345 &  0.0019 &    5\\
2316.6882 & 11.5028 &  0.0009 &   46\\
2316.7888 & 11.5104 &  0.0006 &   53\\
2316.8887 & 11.5079 &  0.0007 &   45\\
2316.9590 & 11.5038 &  0.0011 &   23\\
2352.6743 & 11.5638 &  0.0006 &   47\\
2352.7751 & 11.5613 &  0.0005 &   50\\
2352.8530 & 11.5622 &  0.0006 &   34\\
\enddata
\label{s1040tab}
\tablenotetext{a}{mJD = HJD - 2450000}
\end{deluxetable}

\begin{deluxetable}{cccr}
\tablewidth{0pt}
\scriptsize
\tablecaption{Averaged Photometry for S1063}
\tablehead{\colhead{mJD\tablenotemark{a}}&
\colhead{$\bar{V}$}& \colhead{$\sigma_{\bar{V}}$} & \colhead{$N_{obs}$}}
\startdata
1884.9297 & 13.4520 & 0.0010 &  33\\
1885.0391 & 13.4550 & 0.0014 &  27\\
1886.8594 & 13.4854 & 0.0011 &  36\\
1886.9609 & 13.4899 & 0.0008 &  43\\
1887.0469 & 13.4912 & 0.0010 &  32\\
1890.8711 & 13.5676 & 0.0011 &  43\\
1890.9375 & 13.5754 & 0.0025 &  13\\
1891.8750 & 13.5556 & 0.0012 &  37\\
1891.9883 & 13.5590 & 0.0010 &  52\\
1892.0625 & 13.5538 & 0.0015 &  27\\
1933.7305 & 13.5066 & 0.0007 &  55\\
1933.8320 & 13.5129 & 0.0006 &  64\\
1933.9336 & 13.5140 & 0.0007 &  58\\
1934.0117 & 13.5138 & 0.0010 &  29\\
1935.7305 & 13.5668 & 0.0008 &  45\\
1935.8320 & 13.5704 & 0.0007 &  60\\
1935.9336 & 13.5697 & 0.0007 &  54\\
1936.0000 & 13.5757 & 0.0013 &  17\\
1939.8750 & 13.6060 & 0.0010 &  36\\
1939.9375 & 13.6033 & 0.0027 &   7\\
1940.7070 & 13.5926 & 0.0009 &  37\\
1940.8398 & 13.5944 & 0.0010 &  47\\
1940.9375 & 13.5924 & 0.0011 &  46\\
1941.7188 & 13.5715 & 0.0008 &  50\\
1941.8242 & 13.5696 & 0.0007 &  49\\
1941.9258 & 13.5693 & 0.0007 &  49\\
1941.9922 & 13.5667 & 0.0016 &  12\\
1958.6797 & 13.5500 & 0.0104 &   4\\
1958.8984 & 13.5577 & 0.0008 &  49\\
1958.9531 & 13.5658 & 0.0025 &   5\\
1959.8984 & 13.5886 & 0.0010 &  51\\
1970.6875 & 13.5418 & 0.0008 &  63\\
1970.7930 & 13.5450 & 0.0008 &  75\\
1970.8828 & 13.5474 & 0.0009 &  67\\
1972.6914 & 13.5598 & 0.0007 &  97\\
1972.7930 & 13.5630 & 0.0010 &  89\\
1972.8594 & 13.5639 & 0.0022 &  17\\
1974.6758 & 13.5433 & 0.0024 &  30\\
2296.7188 & 13.5082 & 0.0008 &  21\\
2300.0000 & 13.5386 & 0.0009 &  20\\
2311.6992 & 13.6108 & 0.0010 &  24\\
2311.8008 & 13.6122 & 0.0006 &  44\\
2311.9023 & 13.6116 & 0.0006 &  48\\
2311.9531 & 13.6154 & 0.0027 &   3\\
2316.6875 & 13.5574 & 0.0016 &  42\\
2316.7891 & 13.5538 & 0.0008 &  53\\
2316.8906 & 13.5541 & 0.0009 &  46\\
2316.9609 & 13.5599 & 0.0019 &  22\\
2352.6758 & 13.5144 & 0.0007 &  48\\
2352.7773 & 13.5164 & 0.0006 &  51\\
2352.8555 & 13.5183 & 0.0008 &  32\\
\enddata
\label{s1063tab}
\tablenotetext{a}{mJD = HJD - 2450000}
\end{deluxetable}

\end{document}